Lightning generation in moist convective clouds and constraints on the water abundance in Jupiter

Yury S. Aglyamov and Jonathan Lunine

Department of Astronomy and Carl Sagan Institute, Cornell University

Heidi N. Becker

Jet Propulsion Laboratory, California Institute of Technology, Pasadena, CA

Tristan Guillot

Université Côte d'Azur, Lagrange, Observatoire de la Côte d'Azur, CNRS, Nice, France

Seran G. Gibbard

Livermore National Laboratories, Livermore, CA

Sushil Atreya

Climate and Space Sciences and Engineering, University of Michigan, Ann Arbor MI

Scott J. Bolton

Southwest Research Institute, San Antonio TX

Steven Levin and Shannon T. Brown

NASA Jet Propulsion Laboratory, Pasadena, CA

Michael H. Wong

SETI Institute, Mountain View, CA

Key points

- A model of the generation of lightning is applied to Jupiter, constrained by Juno data; the model reproduces well terrestrial rates
- The model allows for lightning at moderately subsolar abundances of water if ammonia is present.
- The model produces lightning at shallow altitudes (heights above 2 bars) while also producing lightning at deeper levels seen by Galileo.

Abstract

Recent Juno observations have greatly extended the temporal and spatial coverage of lightning detection on Jupiter. We use these data to constrain a model of moist convection and lightning generation in Jupiter's atmosphere, and derive a roughly solar abundance of water at the base of the water cloud. Shallow lightning, observed by Juno (Becker et al., 2020, *Nature*, **584**, 55-58) and defined as flashes originating at altitudes corresponding to pressure less than 2 bars, is reproduced, as is lightning at a deeper range of pressures, including those below the water cloud base. It is found that the generation of lightning requires ammonia to stabilize liquid water at altitudes corresponding to sub-freezing temperatures. We find a range of local water abundances in which lightning is possible, including subsolar values of water--consistent with other determinations of deep oxygen abundance.

Plain Language Summary

Many missions to Jupiter have detected lightning in its atmosphere, but the Juno mission now in orbit has made the most extensive observations on where, and the rate at which, lightning flashes occur. Lightning on Earth is the discharge of electricity that happens when positive and negative charges separate between raindrops and small ice particles in towering thunderheads, somewhat like how combing a cat's fur can create a static electric discharge. This same process is thought to occur in Jupiter's atmosphere, though with some important differences, and we model that process here. Comparing our model with the Juno data also supports the idea that ammonia is in Jupiter's thunderheads, dissolving some of the water ice and increasing the abundance of raindrops.

1. Introduction

Lightning on Jupiter was discovered by Voyager 1 optical images (Cook *et al*. 1979) and plasma wave signatures (Gurnett *et al*. 1979). Subsequently, all spacecraft that have orbited, flown past, or dropped into Jupiter have detected lightning (Lanzerotti *et al*. 1996, Rinnert et al. 1998, Brown *et al*. 2018). Voyager optical and radio data allowed lightning flash energies and total energy dissipation rates to be derived (Borucki *et al*. 1982), and the depth of lightning to be estimated (Borucki and Williams 1986). Subsequent modeling efforts (Yair *et al*. 1995, Gibbard *et al*. 1995) focused on charge separation mechanisms and established that the bulk of the lightning observed in Jupiter likely comes from the water cloud which underlays the visible ammonia clouds.

The Juno mission has provided a leap forward in the general understanding of the distribution, occurrence, and depth of lightning. Detection of lightning at 600 MHz frequency using Juno's microwave radiometer (MWR) reveals lightning to be more frequent near the poles than the equator (Brown et al. 2018). Imaging data from the Stellar Reference Unit (SRU) indicates lightning to be present not only near the 4-6 bar depths inferred from earlier lightning observations (e.g., Borucki and Williams 1986) but also as high in altitude as the 1-2 bar pressure regime (Becker *et al.* 2020).

Interest in Jovian lightning is rooted not only in the intriguing phenomenology of flashes themselves, nor simply as a tracer of moist convective activity (Ingersoll *et al*. 2000), but also in constraining the deep Jovian water abundance (Wong et al. 2008). Lightning on Jupiter is a phenomenon associated with charge separation, and vertical motions in an atmosphere combined with condensation of cloud particles provide the dominant mechanism for efficiently separating charge (Gibbard 1996). Therefore, the rate of occurrence of lightning provides a window into the amount of condensable gas in a moist convective regime. As we show below, if the generation of charge separation involves two phases, liquid and solid, then the depth and flash rates provide a broad constraint on the amount of water in Jupiter's atmosphere.

Charge separation involving liquid and solid phases is the most common mechanism for lightning generation in terrestrial water clouds (Saunders, 2008). We do not attempt to model lightning generation where the entirety of the cloud / convective system is in the solid region, even though the laboratory data we use for the charging model comes from graupel. Our assumption is that a size gradient in the condensates in the cloud is much easier to form with liquid particles than ice particles, consistent with the relative lack of polar lightning on Earth (Christian et al., 2003). Anvil lightning, also cited as an example of lightning in the presence only of a solid phase (Dye and Bansemer, 2019), is created in the wind-sheared regions of a cloud that extends well down into the atmosphere where liquid water is present; in other words, while it happens outside the entraining plume itself, it still depends on it. The depth distribution of Jovian lightning flashes from the spacecraft data we use indicates that charge separation is occurring where two condensable phases are present. The possibility of ice/ice collisions generating lightning (Saunders, 1993) under extreme dynamical conditions only mildly affects the strength of our overarching conclusion that shallow flashes imply the presence of liquid at temperatures well below the pure water freezing point in Jupiter's atmosphere, as discussed in section 6.

For any liquid water condensate to occur in Jupiter's atmosphere in the absence of melting point depression from ammonia, a bulk water abundance (below the water cloud base) that is roughly solar is required. Thus the observation of frequent lightning combined with the most plausible mechanism for lightning generation in moist convective plumes has been used to argue that the bulk water abundance is at least solar (Brown *et al*. 2018), and therefore much larger than the subsolar value measured at the entry site of the Galileo probe (Wong *et al*. 2004).

In this paper we go a step further to constrain the range of allowable water abundances from the Galileo Orbiter and Juno SRU lightning data, by applying models of moist convection in water clouds and a charging model originally developed by Gibbard (1996). Jupiter's atmosphere has been simulated by a number of time-dependent 3-D models of moist convection driven by water (Yair *et al*., 1995a; Palotai *et al*. 2014, Li and Chen, 2019). These models provide important information on the nature of moist convective storm formation in Jupiter. We employ a simpler one-dimensional model in the present paper, because we are specifically interested in exploring the relationship between lightning generation and microphysical processes in Jupiter motivated by recently published data on night-side lightning frequency vs altitude from Juno Stellar Reference Unit (SRU) data (Becker et al., 2020). Our 1-D model is well matched to the data available from the SRU observations, for which there is very limited information on cloud morphology that would be important in a 3-D model. The model we use here also reproduces well terrestrial flash rates with reasonable choices of microphysical parameters.

Our results are intriguing in terms of highlighting the role of ammonia in generating lightning at altitudes above the pure liquid water clouds (so-called "shallow lightning), by providing a liquid water phase at temperatures well below the normal freezing point, or even that of typically supercooled water droplets. We also find that the observed number of flashes provides evidence against excessively large or small water abundances relative to solar. Therefore, while 1-D models cannot be the last word in the study of lightning on Jupiter, they will motivate more detailed future studies of the ammonia-water thunderstorms with higher-dimensional models.

Section 2 describes the numerical model and the improved microphysics over the previously published 1-D model of Gibbard et al (1995), while section 3 describes the choice of parameters. Sections 4 describes the results for both Earth and Jupiter. Section 5 compares the results with available data on terrestrial flash rates, and for Jupiter data from Galileo and Juno. In section 6 we compare our results on the role of ammonia in lightning generation with that of the model of Guillot et al (2020) for formation of large ammonia-water "mushballs", motivated by Juno microwave radiometry data.

2. Numerical model

We built a one-dimensional model of lightning generation based on Gibbard (1996), which includes an entraining plume scheme originally proposed by Stoker (1986), and a charging model described in Gibbard et al. (1997). The only dimension considered is height, with a plume (assumed to be stable in time and of fixed radial extent) carrying parcels of moist air (in Jupiter's case, hydrogen and helium, assumed to behave as an ideal gas) from the level where they become buoyant upwards. As the plume rises, water vapor condenses into particles, which grow by colliding with one another until they become large enough to precipitate out; meanwhile, the plume entrains surrounding dry, stationary air, becoming drier and slower. Droplets of liquid water and ice of different sizes colliding with one another leads to negative charge preferentially moving to larger particles. Since the large particles fall faster, this in turn leads to electric field growth and the buildup of potential energy until a breakdown value of the electric field is achieved and leads to a lightning discharge. We test our model by running cases for lightning on Earth, where data are abundant.

We begin with a parcel of moist air at 10 bars for Jupiter and 1 bar for the Earth, surrounded by dry air of the appropriate composition for each body. So long as the moist air is no more buoyant than the dry air around it, both are assumed to follow an adiabatic profile, a dry adiabat except in the case when the moist air has a partial pressure of water exceeding the saturation vapor pressure calculated as per Lowe (1977), in which case water condenses, releasing its latent heat, and the temperature changes with pressure as per a moist adiabat. In calculating the heat capacity of Jovian air, only hydrogen and helium in the observed Jovian ratio (von Zahn et al. 1998) and water vapor (if any) are considered, as the contribution of other jovian gases is negligible.

Once the moist parcel is buoyant, convection is initiated, and the parcel gains a positive upward velocity that initially increases, as well as being able to keep in suspension particles whose terminal velocity is lower than the parcel's upward vertical velocity, allowing condensate to accumulate and particles to grow. The parcel temperature is assumed to change adiabatically, and the faster the plume rises, the faster it will incorporate dry air from the surrounding, downwelling region. A larger entrainment rate is mathematically equivalent to a smaller plume radius, and leads to more dry air from the surrounding environment introduced into the column. The entrained air follows the environmental temperature profile. In the buoyant region it will be denser than the plume despite having a lower molecular weight, making buoyant convection more difficult. Further, positive buoyancy is required to accelerate initially stationary dry air to the plume's upward velocity. The condensate mass decreases buoyancy so long as it remains suspended in the plume, but this is a smaller effect.

Table 1 lists the symbols used. As in (Gibbard 1996), we use an entrainment rate $\varphi \approx 0.2\,(RT)/(P\mu rg)$ as the fraction of mass added per unit pressure. As such, $dw = -w\varphi\,dP$, $df = -f\,\varphi\,dP$, and the concentration of cloud particles in each size bin is likewise decreased; the plume radius increases, but is capped to be no more than a factor of $\sqrt{2}$ greater than the initial radius to leave room for downdrafts. That is, for a dry adiabat with entrainment we take

$$dT/dP = \{RT_v\}/\{\mu P c_P\} - \varphi(T-T_f), \qquad (1)$$

and for a moist adiabat with entrainment

$$dT/dP = \{[dT/dP]_{dry} + [dT/dP]_{dry}(Lf_S\mu)/(RT_v) + (T-T_f)(\mu Lf_S\varphi)/(RT_v) - (Lf_S\varphi)/c_P\} \,/\, \{1 +(L^2 f_S\varepsilon\mu)/(c_P RT^2)\}, \qquad (2)$$

where $T_v$ is the virtual temperature defined as $T(1+f/\varepsilon)/(1+f)$. For an adiabat outside the plume, either below the level of free convection or in the regions of descending dry air, the terms with entrainment rate are zero. Note that, while it is possible that the external environment might not be characterized by a dry adiabat, what we care about is the temperature contrast created by the different temperature profiles within and outside the plume, which affects the buoyancy. Given that the entrainment is determined by the cross-sectional size of the plume, the distribution of which is not known for Jupiter, introducing a temperature profile different from dry adiabatic would simply add another parameter that is at the moment poorly constrained. We felt it simpler to make the most idealized assumption for the tropospheric environmental temperature gradient. However, inclusion of a stratosphere is important as the reversal of the environmental temperature gradient at the tropopause strongly brakes the ascent of buoyant plumes. As the stratospheric temperature profile does not approximate an adiabat, the Jovian stratospheric environmental temperature profile at pressures less than 0.25 bars is assumed to be equal to that found by the Galileo probe, as per Seiff *et al*. 1998, and the Earth stratosphere to the ISA Standard Atmosphere.

Particles are assumed to immediately reach their terminal velocity, given by

$$w_i^2 = (8b\rho g RT)/(3C\mu P). \tag{3}$$

Particles in bins whose terminal velocity exceeds the upward velocity of the plume are precipitated immediately and removed from the plume, and any remaining particles left over in the plume if it reaches the top of the atmosphere are precipitated out at that point.

From the velocity and the hydrostatic pressure gradient, the time it takes the parcel to ascend through the pressure step of 10 pascals can be calculated and used for particle growth calculation. Both liquid water droplets and ice particles in the cloud are treated as being spherical with a constant density of 1000 kg m$^{-3}$ for liquid and 400 kg m$^{-3}$ for ice (graupel). They are divided into 31 bins according to their liquid-equivalent radius (from which their actual radius, mass, and volume immediately follow), with each bin's upper boundary being a radius $\sqrt{2}$ times its lower boundary. The smallest bin stretches from a radius of 10 microns to a radius of 14.1 microns, with the largest bin's upper bound being a radius of approximately 46.3 centimeters (not reached in practice, but estimated to be near the maximum size of Jovian hail). Particles initially condense with a flat distribution of radii in the smallest bin; changing the radius of this bin by a factor of 2 in either direction changes flash rates by less than 10%, demonstrating stability with respect to the choice of size bin. The arithmetic mean of these upper and lower bounds, $b_0$, is used as an approximate radius of particles within the bin in some of the calculations to follow We describe the size distribution of particles by a piecewise linear function F(b) such that the number of particles with radius between $b_1$ and $b_2$ in a region is [the integral of F(b) from $b_1$ to $b_2$, db]*[volume]. As such, F(b) has values with dimension m$^{-4}$. In a given size bin, it is assumed that F(b) is linear except if this would cause it to be negative, $F(b) = \max[0, n_0 + k(b-b_0)]$, defined by two parameters per bin, $n_0$ (with units $m^{-4}$) and $k$ (with units $m^{-5}$); for empty bins $n_0 = k = 0$. In the largest non-empty size bin, we thus tend to have F(b) be piecewise linear within the bin, with a negative slope until F(b)=0 and constantly zero thereafter. It is assumed that all particles within a bin precipitate together, using the terminal velocity of a particle of radius $b_0$. The total volume (and from it the total mass M) of particles in a given bin can thus be found by integrating $(4\pi/3)b^3 F(b)$ from the lower bound of the bin to either its upper bound or the x-intercept of F(b); the average mass can be found by dividing M by the integral of F(b) over that range, which is the number density N.

It is assumed that, whenever two cloud particles collide, they will stick together to form a single particle with probability equal to $E_i$. Calculating the frequency of such collisions requires knowing the typical relative velocities of cloud particles. It is assumed that the velocity is predominantly vertical, with the relative velocity of the cloud particles being the difference between the terminal velocity at the upper bound of the larger bin and the lower bound of the smaller bin. This ensures that particles from the same bin have a nonzero relative velocity and can merge with one another. As such, the number of collisions $n_{coll}(i,j)$ between a particle from bin i and a particle from bin j, $i \geq j$, per unit time and per unit volume is

$$n_{coll}(i,j) = \pi (b_0(i)^2 + b_0(j)^2) w_{rel}(i,j) N(i) N(j) E_i. \qquad (4)$$

Multiplying by the time step gives a number of collisions per unit volume.

Collisions are assumed to remove a number of particles from the smaller bin equal to the number of collisions, with average mass equal to the average mass of the smaller bin. This mass is in its entirety transferred to the larger bin. However, this increase will cause some of the particles in the larger bin to be transferred from bin i to bin i+1. For this calculation, it is necessary to make simplifying assumptions. We define the radius $b_x$ as the radius of a particle that, when colliding with a particle of radius $b_0(j)$, would produce a particle with radius equal to the upper bound of bin i. Then, it is assumed that exactly the particles with radius greater than $b_x$ are transferred to bin i+1 upon collision, so that their total mass and number are transferred up a bin, along with the proportionate amount of mass initially from bin j. In this fashion, the total mass and number density are updated for each bin. Following this, $n_0$ and $k$ are rederived from M and N, with the formula dependent on whether or not F(b) has an x-intercept within the bin, and particle growth is iterated for the duration of the ascent through the pressure step.

In an actual storm, there would be a gradient between fully liquid and fully frozen particles. Here, to make the calculation tenable, we assume that all particles at a given altitude are either liquid water or ice; the boundary between them is set at a specific temperature, as the freezing point of water does not change significantly with pressure in this pressure range. This freezing point is set at 273 K or lower, the latter to simulate the possible supercooling of water—a complex phenomenon involving multiple microphysical effects—down to 233K (e.g. Pruppacher & Klett 1997), or a more important depression (possible down to 173K) due to the dilution of ammonia into water (Guillot et al, 2020). The difference between liquid water and ice lies in the values of $E_i$ and $E_q$, the charging efficiencies. These efficiencies hold different values for liquid and ice: that is, $E_i(liquid) + E_q(liquid) = 1$, and $E_i(solid) + E_q(solid) = 1$. Since ice is assumed to satisfy $E_i=0$ and thus $E_q=1$, as $E_q(liquid)$ decreases, charging and lightning will increasingly be dominated by that occurring at temperatures below freezing and therefore shift upward. However this also implies that $E_i(liquid)$ –a variable parameter-- increases, allowing for faster growth of particle size contrast and more lightning overall.

In this fashion, at each pressure level, a particle size distribution is calculated, as well as the sum of the size distribution of precipitation coming from above. The rest of the computations to determine electric field buildup do not alter this distribution; hence, electrostatic levitation (which might serve to increase the mean particle size at which fallout occurs) is not included. Furthermore, particle evaporation (virga) at altitudes above the 10-bar level and particle re-lofting (hail) are not considered. The following calculations are performed at each pressure level, but using a step size of 1 kilopascal.

Firstly, the precipitation coming from above is weighted by the fraction of time it spends within the range of the pressure step. The height of this step is calculated as before, and the particle fall velocity as terminal velocity minus plume velocity. Any precipitating particles with a terminal velocity slower than the plume velocity are considered uninvolved in lightning production (in practice they would either be re-lifted or precipitate outside the plume area), while the upward-moving plume particles are added later. For a set particle radius, the time spent in a certain height range would be equal to height over velocity, but this leads to infinite values for particles just large enough to have a terminal velocity equal to the plume velocity. As such, each contributing particle radius within the bin is also weighted by the difference between its terminal velocity and the updraft velocity. The natural (asymptotic) weighting is $f_{bin} = w_W/(w_i-w)$, where $w_W$ is the updraft velocity at the point where the particle was dropped by the plume. Instead, we perform the calculation for particles slightly larger than the critical radius by treating the bin as if its lower bound was at the critical radius and taking an average velocity for this shrunken bin, leading to a weighting $f_{bin} = w_W [b_+ - b_m] / [\int w_i(b)-w \, db]$, where the integration is done from $b_m$ to $b_m\sqrt{2}$. This yields:

$$f_{bin} = w_W [3 b_+ - 3 b_m]/[b_m [3 w - 2 w_i(b_m)] - \sqrt{2} b_m [3 w - 2 w_i(\sqrt{2} b_m)]], \qquad (5)$$

where $w_W$ is the plume velocity at the altitude of the precipitation's origin, $w$ the plume velocity at the altitude of the electric field growth, $b_+$ is the bin's upper bound, and $b_m$ is the greater of the bin's lower bound and the critical levitating radius. The factor of $\sqrt{2}$ is the factor between the bin boundaries, as near the singularities $b_m$ approaches $b_+$ and we wish to lower the weight of such bins to reflect that most particles in them are being lofted in these conditions.

This creates a weighted precipitation size distribution. Because $n_0$ and $k$ do not in general add linearly, we convert from $n_0$ and $k$ to $M$ and $N$, add the weighted precipitation size distribution to the suspended cloud particle size distribution, and generate an overall particle size distribution that is then converted back to $n_0$ and $k$. We assume this population is stable in time; in reality particles fall down and similar particles, which may have a charging history involving different conditions, fall from above.

Non-inductive charge transfer is assumed, with the empirically observed formula for the amount of charge transferred between colliding ice particles of Keith and Saunders (1989). It is assumed that a collision between two particles will lead to charge transfer with probability $E_q=1-E_i$. As a result, particles of equal size colliding will have no net effect. As such, $w_{rel}(i,j)$ is calculated simply as the difference between the terminal velocities $w_i(b_0(i))$ and $w_i(b_0(j))$, the arithmetic means of the bin bounds, and collision probability is calculated as with particle growth. In this fashion, a rate of charge accumulation per particle is calculated for each bin; under the assumptions being made, this rate is constant in time. Since particles are therefore charged linearly, the electric current caused by faster fall velocities of larger particles also increases linearly in time, and therefore the electric field is quadratic in time,

$$E = \sum[N(i) \, dQ(i)/dt \, [w_i(b_0(i)) - w] \, /\varepsilon_0] \, t_L^2/2. \qquad (6)$$

The current density is linear in time:

$$J = \sum[N(i)\, dQ(i)/dt\, [w_i(b_0(i)) - w]]\, t_L. \qquad (7)$$

When the electric field reaches its breakdown value, a lightning discharge occurs. This breakdown value is taken as $(5\ m^2\ C^{-1})P$ for hydrogen on Jupiter and $(3\ m^2\ C^{-1})P$ for nitrogen/oxygen on Earth from Gibbard (1996), based on a combination of theoretical modeling and observational data. The energy released per unit volume is equal to this breakdown field, times the total charge transferred per unit area. When lightning thus occurs, the charge is assumed to reset to zero at that location and begins building up as before. Thus, dividing the energy per volume by the time-to-discharge yields a rate of time-averaged energy production per unit volume. Integrating over pressure leads to a power per unit area. The calculated power is a rate per unit area of updraft; that is, it does not include downwelling regions of the storm, and it does not include sections of the atmosphere that do not contain a thunderstorm at a given time.

While the model as described in equations (1)-(7) is based on that in Gibbard (1996), there are substantial differences, the greatest being a different assumption of equilibration. In the Gibbard model, convection is modeled separately from particle growth and charging; as such, a plume is modeled without considering precipitation, and then particle growth and charging is assumed to happen in a parcel of air without new influx of material. Since this would ultimately result in all particles growing large enough to precipitate, the particle growth must be limited to an arbitrary time interval; this provides an additional tunable parameter, but one that is artificial. We instead assume that water or ice particles grow in a parcel of air as it ascends, precipitating out once their terminal velocity exceeds the upward velocity of the plume, with the implicit assumption being that the overall vertical structure of the plume is stable in time, with air and water circulating through it. While the plume structure evolves during a thunderstorm, in broad terms (which a constant entrainment rate implies) this is a reasonable approximation: with model velocities it takes 1-2 hours for the plume to ascend to the stratosphere, which is short compared to storm lifetimes.

The calculation for non-inductive particle charging was, as in Gibbard (1996), done separately, following the assumption of a stable plume. However, because we calculate the particle size distribution at each altitude, we are able to include both precipitation and cloud particles in calculating particle charging. We also included a phase change, altered the particle growth algorithm, particularly in calculating collision velocities, and targeted the calculation towards obtaining a flash rate per unit area rather than a time to first flash.

3. Choice of parameters

The model involves a number of known parameters, such as the molecular mass of Jovian air and the heat capacities of the substances involved, the latter of which were assumed to be the NIST-JANAF values. Furthermore, values such as the size of each particle radius bin or the time step should not affect the calculation so long as they are sufficiently small. However, seven parameters represent genuine unknowns about the atmosphere of Jupiter.

The water abundance in Jupiter's atmosphere, below its clouds, is not well known due to spatial variation and model dependencies. Given the solar composition as per Asplund *et al*. (2009), if all oxygen were in the form of water, the water mass fraction in Jupiter's atmosphere would be approximately 0.00625; this is referred to as 1-solar water abundance. The Galileo probe measured a subsolar oxygen abundance, despite most other measured elements being enriched by a factor of 2-3 relative to the solar atmosphere, but this may have been an atypical value due to dynamical effects leading to subsidence and drying in the "5-micron hot spot" into which the probe fell (Orton et al. 1998, Atreya et al. 1999). Juno's MWR has found a deep-water value at a latitude just north of the equator of between one and five times solar (Li et al., 2020). Within the Great Red Spot, high-resolution 5-µm spectroscopy constrains water to the 1.1-12x solar range (Bjoraker et al. 2018). A greater water concentration leads to earlier and more extensive water condensation, and therefore greater moist convection and more lightning, although other parameters such as the temperature profile are also known to have an effect (Yair et al. 1998). Since we as yet do not know the deep water abundance for regions far from the equator, four values considered for water mass fraction were 0.3-solar, 1-solar, 3-solar, and 10-solar, spanning the wide range of values consistent with the Galileo Probe measurement, modeling of the disequilibrium CO abundance, and tropospheric gravity waves radiating from Shoemaker-Levy 9 impacts (Wong et al. 2004, Wang et al. 2015, Ingersoll and Kanamori 1995). Given the number of other parameters that are uncertain, finer water abundance increments are not useful.

The temperature of Jupiter's atmosphere at 10 bars pressure was measured at 338 Kelvin by the Galileo probe (Seiff *et al*. 1998). Because the probe descended in a 5-micron hot spot atypical of most of Jupiter, we consider possible temperature values of either 330 K or 350 K, with most calculations done at the cooler temperature. This may be larger than the plausible range of 10-bar temperatures, but bounds the problem. Higher temperatures lead to water condensing higher up in the atmosphere, but also higher updraft velocities, with differing signs of effect on lightning; overall, the temperature profiles lead to similar conclusions. It is possible to have the plume begin at a higher temperature than its surroundings, but this is not necessary for convection because the difference between moist and dry adiabats creates a temperature contrast between the plume and its surroundings with increasing altitude (under our assumption of a dry adiabatic environmental temperature structure).

The drag coefficient observed in Earth rain varies depending on raindrop size, taking its lowest value at a drop radius near 1 mm and rising for both smaller and larger sizes (Spilhaus 1948). Additional complications are added by the state of water, and the history of that state. We assume the drag coefficient of a sphere in turbulent flow, which is approximately 0.5, for all cloud and precipitation particles. Higher drag coefficients correspond to lower terminal velocities. While this enables larger particles to be supported by the plume, its main effect is to greatly slow particle growth and charging due to lower collision velocities, thus leading to lower flash rates.

The collision efficiency and the charging efficiency both affect the generation of lightning, and which of them is the greater constraint depends on conditions. Their actual values depend enormously on the physical state (shape, stickiness, etc) of the condensate particles being considered, which for Jupiter are unknown even in general. We optimistically assume that their sum is unity, that is, that every collision not resulting in particles merging transfers charge. Furthermore, we assume that ice has a collision efficiency of zero (and so a charging efficiency of unity), because ice particles are unlikely to stick upon collision; this explains the need for liquid water in lightning-generating systems. Then, we find that the model best fits observed behavior of Earth's atmosphere if liquid water has $E_i$=0.8 (and so $E_q$=0.2); see Table 2 and figure 1. Jovian calculations for $E_i$=0.5 ($E_q$=0.5) were also performed, and found to be qualitatively similar. A complete absence of charging in liquids would imply $E_i$=1 ($E_q$=0), but some liquid electrification is observed on Earth (Stetten et *al*. 2019); therefore, a small but nonzero charging efficiency is a reasonable result.

The radius of the entraining plume is set by the properties of convection. An overly small plume will entrain so much material that it stifles itself, whereas a very large plume will break up into smaller plumes (figure 1). In the model, the latter process does not happen, and so it is necessary to choose a plume size that is big enough to be convective but small enough not to escape the atmosphere. For Jupiter, a plume radius of 5 kilometers fits these conditions and is consistent with the results in Yair et al., (1995a). Hueso et al (2002) consider storms of 25 km, but refer to this as an upper limit beyond which their storms break up. Since storms may consist of multiple upwelling plumes, our choice is not necessarily inconsistent with their upper limit. For Earth, typical thunderstorms have upwelling cores that are roughly a kilometer in diameter, though larger upwellings also occur (Stolzenberg, 1998).

The energy of a single flash has a strictly inverse linear effect on the flash rate: the lightning power per unit area does not in this model depend on the flash energy, and thus, for example, doubling the flash energy will always halve the flash rate. As such, the flash rate can be calculated for multiple values of flash energy without rerunning the model, and we use constant values of flash energy for both settings: Given typical total flash energies of $1.5*10^9$ J as per Maggio et al. (2008), the peak observed rate in updrafts corresponds to an electrical energy flux of 2 W/m$^2$; with the chosen parameters for collision efficiency, our model matches this well, predicting maximum values 0.1 to 4 W/m$^2$ depending on temperature. Less intense lightning flashes on Jupiter, as observed by Juno, would correspond to higher predicted flash rates. These values refer to the total electrical energy released by the flash, and not to the optical flash energy, which may be several orders of magnitude smaller.

4. Results for the Earth and for Jupiter

To calibrate the model, Earth cases were considered with a surface (sea level) temperature of 280, 290, 300 or 310 Kelvin, 90% humidity, and environmental temperature 5.5 to 14.5 K higher than plume temperature at sea level (to prevent immediate buoyancy of the vapor-containing plume), for various collision efficiencies with a plume radius of 1 kilometer, consistent with observations of terrestrial thunderstorms. Under these conditions, a buoyant plume is formed, with maximum upward velocities between 15 and 40 meters per second (figure 1).

When considering the flash rates of lightning to the depth of 10 bars in Jovian conditions, the flash rate is mainly determined by the water abundance and the freezing point (tables 3 through 8 and figures 2-7). With insufficient freezing point depression, no lightning occurs; otherwise, the flash rate rapidly climbs to a value largely dependent on water abundance alone. This value ranges from about 200 W/m$^2$ for 0.3x solar water to 15 000 W/m$^2$ for 10x solar water, and the minimal amount of freezing point depression drops from 60 K to 20 K over that range.

Lightning occurs over a wide range of pressure levels. A substantial fraction of lightning occurs in the 0-2 bar region, where even ammonia-containing water is likely to solidify (tables 7 and 8). However, when lightning occurs in the local absence of liquid, it then relies on advection of material from below, as particle growth does not take place without a liquid phase. For a freezing point above 253 K and 3x solar or lower water, no lightning at all is observed, due to the absence of liquid above the level of free convection. For parameter values near this boundary, electrical power is extremely low, and its value and distribution change very rapidly compared to the size of the parameter grid.

The electrical power rises slightly faster than linearly with water abundance, if the freezing point is sufficiently low. With greater water abundance, convection shifts deeper into the atmosphere, requiring less freezing point depression; but due to greater convective vigor, lightning rates increase throughout the atmosphere, and the depth distribution of available electrical power does not greatly change, though medium and deep lightning rates grow faster than shallow lightning rates. Similarly, if the freezing point is not far below 273 K, a colder temperature profile leads to deeper condensation and greater lightning rates.

An extreme freezing point depression of more than 60 K—much greater than the 0-40K drop generally inferred for terrestrial clouds where $NH_3$ is absent-- slightly decreases lightning rates, as the condensate being liquid throughout the plume lowers the efficiency of electric charging at high levels, and little particle growth occurs so high up regardless. At these extremely low freezing points, the effect of 10-bar temperature on lightning rate lessens and in some cases reverses, because the colder profiles have less vigorous (albeit deeper) convection.

5. Comparison with observations

For mesoscale convective complexes on Earth, the peak rate of cloud-to-ground discharges has been observed to be on the order of $1.5*10^{-3}$ flashes per square kilometer per second, which must be doubled when restricted to only updrafts; for the system as a whole, though, the maximum rate was observed at $3.3*10^{-6}$ flashes per square kilometer per second (Goodman and MacGorman 1986). For the entire Earth, including regions without cloud cover, the often-quoted value is 100 flashes per second, corresponding to a flash rate of $2*10^{-7}$ flashes per square kilometer per second (Orville and Spencer 1979). More recent space-based observations have given values a factor of two lower (Christian et al., 2003). Given typical total flash energies of $1.5*10^9$ J, we find that our model is consistent within a factor of a few with peak flash rates observed for Earth within convective storms.

The rate and altitude of lightning flashes and temperature dependence for the model, as applied to Earth conditions, best fits observations for a water collision efficiency of 0.8 and a freezing point of 253 K. The former constraint is relatively weak, with collision efficiencies of 0.5 or 0.9 producing similar results. A collision efficiency of 1.0, however, produces lightning concentrated at overly low pressures. Flash rates peak near the 0.5-bar level, higher at lower freezing points; however, flashes are observed at a variety of altitudes, down to the ground. While the model does not specifically simulate cloud-to-ground flashes, flashes in the lowermost atmosphere represent charge transfer to the ground, which leads to cloud-to-ground flashes in practice. Greater freezing point depression causes flashes to increasingly occur in the stratosphere, and also greatly decreases the dependence of flash rate on temperature; for comparable levels of free convection, supercooling of 10-20 Kelvin leads to the flash rate varying by a factor of 4-10 per 10 K base temperature, consistent with observation. Observationally, some supercooled water droplets have been observed at temperatures as low as 233 K, but a lower upper limit is reasonable for the bulk. This offers a constraint for interpreting the Jovian results: freezing point depression greater than ~20 K cannot, within the model, be explained by supercooled water.

Galileo's observations of Jupiter's global flash rate were on the order of 10 strokes per second (Little et al, 1999), less than Earth over an area 100 times greater; Juno observed a higher flash rate but much optically dimmer flashes summing to a lower energy flux, implying that a different part of the lightning distribution was being sampled. This sets an absolute lower bound of $10^{-4}$ watts per square meter for the power flux assuming a flash energy of $10^{12}$J as per Yair et al. (2008, 1995a). For individual storms, an areal lightning analysis for Galileo for three storms found a flash rate of 0.2 s$^{-1}$ for each on average (Little *et al*. 1999). The sizes of the storms have large uncertainties ranging between $10^5$ and $1.4*10^6$ km$^2$.The implied flash rate is thus between $1.4*10^{-7}$ and $2.0*10^{-6}$ s$^{-1}$ km$^{-2}$, and likely closer to the lower part of this range. This rate is multiplied by four to account for the storm's non-rectangularity and the portion of it taken up by downdrafts. This provides a lower bound between 0.9 and 12 W m$^{-2}$ for flash energies as per Yair et al. (2008), far lower than the bulk of model predictions. The reason is that the model is designed to predict a peak flash rate, which requires higher resolution that can be achieved for Jupiter. However, the prevalence of lightning is sufficient to exclude parameter combinations that produce no lightning at all and provide additional evidence against 0.1x solar water.

The depth of lightning observed on Jupiter can be estimated, via a model of atmospheric scattering, from the half-width at half-maximum (HWHM) of the flash's spatial extent. Such estimates have been reached for several flashes by the Galileo Solid State Imager and the Juno Stellar Reference Unit (Little *et al*. 1999, Dyudina et al. 2002, Becker *et al*. 2020). Estimates were also made from Voyager 1 observations, but given the lower spatial/temporal resolution, doubts have been raised as to whether some of these recorded flashes were in reality overlapping sets of smaller flashes. While the Juno observed flash rate was 15 times higher than the Galileo flash rate, average Galileo flash energies were ~150 times higher, and measured at greater depths. The sign of this difference is expected: Galileo's measurements were made with longer exposure time but lower sensitivity than the Juno SRU, meaning that fewer but more intense lightning flashes should be observed, and the breakdown field increases with pressure, explaining the higher intensity in deep flashes. In SRU data the most intense Juno flashes were almost equal in energy to the weakest Galileo flashes, implying that the two energy ranges can be concatenated to find the overall rate; however, if very infrequent flashes more intense than the Galileo range were found to occur on Jupiter, these conclusions would be altered.

Quantitative differences suffer from small-number statistics. The Juno SRU observed four flashes with HWHM of 40 km or less, corresponding to a pressure level of 0-2 bars (hereafter shallow flashes), and two with HWHM between 40 and ~ 55 km, corresponding to a pressure level between 2 and 4 bars (hereafter medium-depth flashes). The latter were on average more energetic, with 55% of the energy coming from medium-depth flashes and 45% from shallow flashes. The Galileo observed energy flux (flashes per unit area times the average flash energy) was ten times higher, but only three flashes had depths measured, one (25% of the energy) a medium-depth flash, and two with half-widths above 60 km, expected to correspond to pressure levels of 4-10 bars (hereafter deep flashes). Dyudina et al. (2002) rederived depths and added more flashes, generally consistent with the Little et al. (1999) conclusion that Galileo observed deep flashes. Some of these flashes might occur below the cloud base; this is expected, as precipitation should carry charge differential down below cloud level, until the depth at which it evaporates. The precise depth at which this occurs is unknown and is unlikely to be exactly the 10 bars used in the model, but flash rates below the cloud base do not greatly increase with depth, so the effect of this uncertainty is limited.

As such, observations show that approximately 30% of flash energy is released at 0-4 bars, and that among those 0-4 bar flashes, 13% of the energy is released at 0-2 bars. This is reached based on a very small number of observations, but is most consistent with the model results for a freezing point near 213 K. For lightning at 0-4 bars pressure to output a minority of the energy, a minimum 40 K of freezing point depression - greater than the best fit for Earth - is required. While no clear conclusion on water abundance can be reached, 0.1x or 0.3x solar water require 80K or greater freezing point depression to satisfy this constraint for either temperature profile. The upper limit on water abundance is global, as lightning will not preferentially form in the absence of clouds; the lower limit, by the same reasoning, is local, as lightning may only occur in limited wet regions of Jupiter. Further constraints would require continued measurements of the depth distribution of Jovian lightning, or else information on local flash rates.

6. Interpretation

The observation of lightning at altitudes above the 2-bar pressure level by Becker *et al*. (2020) is fully consistent with the model. Lightning power at these shallow levels is estimated to be 10-100 times smaller than the total lightning power down to 10 bars; however, as conditions in this region are more Earthlike, it is expected that such flashes would be lower-energy and therefore more common. This is precisely what we see in the combined Juno and Galileo data. While the water cloud base is far below this level, the water cloud tops are substantially higher, and lower breakdown fields are required to generate lightning at higher altitudes. The predicted distribution is smooth rather than bimodal, but a bimodal distribution seems unlikely unless lightning is being generated in the pure ammonia clouds (which are above the 0.7-bar level), a situation that is unlikely given that the condensate in those clouds is low-density solid. Both observation and model thus imply that lightning at Jupiter occurs in a broad altitude range, rather than at a singular pressure level.

Conversely, the observed shallow lightning seems to call for liquid water to occur at altitudes where the temperature is so far below the pure water freezing point that even supercooling cannot be invoked. Therefore, freezing point depression such as that caused by ammonia solvation in the liquid seems a plausible mechanism. But whether ammonia is necessary to stabilize liquid water for lightning generation depends chiefly on the level of free convection. By the nature of both the model and the real Jovian atmosphere, particle growth and charging are both required for lightning generation, but they need not happen at the same time. If the plume rises from a level where liquid water is stable, allowing particle growth, ice particles will still charge one another in the upper reaches of the atmosphere to generate electric fields without the need for ammonia at the 1-2 bar level to keep the water liquid. However, if it is assumed that ice particles are 'non-sticky' and do not grow by collisions (or grow at very low efficiency), then this requires very vigorous convection ongoing below the ice line. While the base of the water cloud (the lifting condensation level) occurs at depths below the ice line for a subset of reasonable Jovian parameters, the level of free convection where the plume becomes buoyant is higher. This gap is further increased for Jupiter, relative to Earth, because water vapor has a lower molecular mass than nitrogen and is therefore buoyant on Earth, whereas in Jupiter's hydrogen-helium atmosphere the compositionally driven buoyancy is negative. In our model, the level of free convection typically occurs near 4 bars. In what follows, we do not consider additional speculative sources of convection such as deep alkali metal moist convection (Lunine *et al*. 1989) or mechanically forced convection for which we have no evidence at this time. A quantitative estimate of the effect this has can be gathered from varying the freezing point. Earth observations, as described above, require 10-20 K supercooling (Rutlege and Hobbs 1984, Pruppacher and Klett 1997). For Jupiter, given a 330 K base temperature, the water freezing point must be lowered, at the warmest, to 253 K for 3-solar water, to 233 K for 1-solar water, or to 213 K for 0.3-solar water to generate observable amounts of lightning; for energetic lightning to be mostly deep, even greater freezing point depression is required. These depressions are at the limits of supercooling in the terrestrial context, implying that water cloud particles on Jupiter have substantial dissolved ammonia. The great latitudinal variability in lightning rates may indicate that Jupiter is in a region of the phase space where lightning rates vary extremely rapidly with changing parameters; this would imply that the true ammonia abundance corresponds to between 20 and 40 Kelvin freezing point depression in addition to supercooling, that is, water cloud droplets that contain 10-25 percent ammonia by weight relative to the total condensate (ammonia plus water). More ammonia is needed for lower deep-water abundances, but even for subsolar water abundance, lightning can be generated if enough ammonia to create liquid is present, and even for supersolar water abundance some ammonia is required.

In order for ammonia to depress the freezing point sufficiently, there must be enough of it relative to water to maintain a significant liquid fraction. In fact, equilibrium thermodynamics predict that, in Jupiter conditions, ammonia and water can mix to form a liquid at pressures between 1.1 and 1.5 bar and temperatures between 173K and 188K, as shown by Guillot *et al*. (2020). The consequent growth of these water-ammonia particles ('mushballs') and their fall through the cloud should lead to the presence of partially liquid particles in a wider range of pressure levels. The distribution of these particles in terms of sizes and pressure levels remains an open question. A model run with a separate liquid region at 173-188 Kelvin in addition to freezing point depression, as would naively follow from Guillot *et al*.'s equilibrium calculations, has been tested, with lightning being generated at lower water abundances than with freezing point depression alone, but the qualitative conclusions about the water abundance remaining the same.  Full analysis of the mushball model will require modeling time-dependent processes in condensate formation, followed ultimately by three-dimensional convective modeling of mixed ammonia-water storms.

7. Conclusions

Using a one-dimensional model of moist convection and lightning generation, we have replicated terrestrial lightning flash rates and altitudes with a reasonable choice of parameters. These, in turn, allow us to use the model to fit the lightning rates and altitudes observed by Galileo and Juno. We find that the presence of relatively abundant shallow lightning (1-2 bars), combined with other observations, requires ammonia to act as an antifreeze for the water droplets.  With a sufficient amount of such dissolved ammonia, subsolar water abundances are consistent with the observation of lightning, and the bulk ratio of the two species (leaving aside variations as condensation progresses) allows for sufficient melting point depression.  Our model predicts lightning over a wide pressure range, rather than a single 'depth of lightning'.  The values for deep water allowed by the model are consistent with suggestions from interior models (Wahl *et al*. 2017) that bulk water might be severely depleted in the deep interior, and with analysis of equatorial MWR data (Li *et al*. 2020). They allow for a wide range of compositions of the planetesimals seeding Jupiter during its formation, with the exception of clathrate formed in a nebula with solar oxygen abundance (Gautier *et al*. 2001).

Acknowledgements

**Table 1** Definition of symbols

| | | Units |
|---|---|---|
| b | radius of a given cloud particle | m |
| $b_0$ | approximate radius of particles in a given bin,= arithmetic mean of its bounds | m |
| C | aerodynamic drag coefficient of a particle, assumed constant | -- |
| $c_P$ | specific heat capacity of dry air, taken as 14.5 kJ K$^{-1}$ kg$^{-1}$ for Jupiter | kJ K$^{-1}$kg$^{-1}$ |
| E | electric field at a given pressure level | N C$^{-1}$ |
| $E_i$ | collision efficiency = probability that two colliding cloud particles will merge | -- |
| $E_q$ | charging efficiency = probability that two colliding particles will transfer charge | -- |
| f | *mass* fraction of condensate *vapor* in air, 0.00625 for solar water abundance on Jupiter | -- |
| $f_S$ | *mass* fraction of condensate *vapor* in the air that corresponds to saturation vapor pressure | -- |
| g | gravitational acceleration, assumed constant, 24.79 m s$^{-2}$ for Jupiter | m s$^{-2}$ |
| J | vertical current density at a given pressure level | C s$^{-1}$ m$^{-2}$ |
| k | slope of the particle density distribution as a function of radius | m$^{-5}$ |
| L | latent heat of vaporization of the condensate (water), assumed constant at 2.257 MJ kg$^{-1}$ | MJ kg$^{-1}$ |
| M | total mass density of cloud particles within the size range of a bin | kg m$^{-3}$ |
| N | total number density of cloud particles within the size range of a bin | m$^{-3}$ |
| $n_0$ | particle density distribution as a function of radius at the center of a bin | m$^{-4}$ |
| P | pressure level under consideration | Pascals |
| R | universal gas constant 8.314 J K$^{-1}$ mol$^{-1}$ | J K$^{-1}$mol$^{-1}$ |
| r | radius of the entraining plume | m |
| T | temperature of the ascending air parcel | K |
| $T_f$ | temperature of the dry air surrounding the plume | K |
| $t_L$ | time since the last lightning discharge | s |
| w | upward velocity of the entraining plume at a given point | m s$^{-1}$ |
| $w_i$ | terminal velocity for a particle of a given size relative to its surroundings | m s$^{-1}$ |
| $w_{rel}$ | average relative velocity of particles in two size bins | m s$^{-1}$ |
| ε | ratio of the molar mass of the condensate (water) to that of dry air, 8.2 for Jupiter. | -- |
| μ | molar mass of dry air, assumed constant, 0.0022 kg mol$^{-1}$ for Jupiter | kg mol$^{-1}$ |
| φ | fraction of the entraining plume replaced with dry air per unit pressure rise | -- |

***Table 2:*** *Earth, electrical energy flux (W m$^{-2}$) vs. Temperature, Liquid Collision Efficiency, and Liquid*

| Freezing point, efficiency | Solution Freezing Point [a] | | | |
|---|---|---|---|---|
| | 280 K, 285.5 K | 290 K, 298.5 K | 300 K, 311.5 K | 310 K, 324.5 K |
| 273K, 0.5 | 0 | 0 | 0.3 | 2.8 |
| 263K, 0.5 | 0 | 0.1 | 1.2 | 4.6 |
| 253K, 0.5 | 0.1 | 0.5 | 2.0 | 5.4 |
| 243K, 0.5 | 0.4 | 0.7 | 2.3 | 5.6 |
| 233K, 0.5 | 0.5 | 0.9 | 2.4 | 5.7 |
| 273K, 0.8 | 0 | 0 | 0.4 | 4.7 |
| 263K, 0.8 | 0 | 0.2 | 2.1 | 7.1 |
| 253K, 0.8 | 0.2 | 0.7 | 3.0 | 7.5 |
| 243K, 0.8 | 0.4 | 1.0 | 3.3 | 7.1 |
| 233K, 0.8 | 0.5 | 1.1 | 3.0 | 7.0 |
| 273K, 1.0 | 0 | 0 | 0.5 | 5.5 |
| 263K, 1.0 | 0 | 0.2 | 2.3 | 6.9 |
| 253K, 1.0 | 0.1 | 0.7 | 2.8 | 5.7 |
| 243K, 1.0 | 0.3 | 0.8 | 2.1 | 3.9 |
| 233K, 1.0 | 0.3 | 0.6 | 1.3 | 2.4 |

[a] *First temperature is plume temperature at 1 bar; second temperature is environment temperature at 1 bar. Initial temperature contrast must be negative on Earth because water vapor has a lower molecular weight than nitrogen and oxygen, and is thus innately buoyant. Values of temperature contrast chosen to ensure level of free convection is at approximately 0.7 bar for all temperatures.*

**Table 3:** *Electrical energy flux above 10*

*Table 3:* *Electrical energy flux above 10 bars (W m$^{-2}$) vs. Water Abundance and Liquid Solution Freezing Point.  Liquid collision efficiency is 0.8, temperature at 10 bars is 330 K.*

| Freezing point | 0.1 x | 0.3x | 1x | 3x | 10x |
|---|---|---|---|---|---|
| 273K | 0 | 0 | 0 | 0 | 3367.6 |
| 253K | 0 | 0 | 0 | 1206.7 | 13384.4 |
| 233K | 0 | 1.0 | 322.3 | 3465.6 | 14045.6 |
| 213K | 2.1 | 59.6 | 854.1 | 4070.3 | 16060.7 |
| 193K | 21.3 | 213.9 | 1170.8 | 4401.4 | 16641.9 |
| 173K | 26.9 | 241.3 | 1268.1 | 4577.7 | 17038.1 |
| 153K | 27.9 | 250.1 | 1269.9 | 4593.3 | 17100.5 |

*Table 4:* *Electrical energy flux above 10 bars (W m$^{-2}$) vs. Water Abundance and Liquid Solution Freezing Point. Liquid collision efficiency is 0.5, temperature at 10 bars is 330 K.*

| Freezing point | 0.1 x | 0.3x | 1x | 3x | 10x |
|---|---|---|---|---|---|
| 273K | 0 | 0 | 0 | 0 | 1988.0 |
| 253K | 0 | 0 | 0 | 754.1 | 9781.8 |
| 233K | 0 | 1.0 | 207.6 | 2061.7 | 14699.9 |
| 213K | 1.9 | 39.3 | 472.6 | 3046.5 | 19049.9 |
| 193K | 17.0 | 110.2 | 1055.0 | 4855.5 | 20393.6 |
| 173K | 29.0 | 211.0 | 1212.5 | 5049.6 | 21190.4 |
| 153K | 32.7 | 220.1 | 1298.9 | 5216.8 | 21250.3 |

**Table 5:** Electrical energy flux above 10 bars (W m⁻²) vs. Water Abundance and Liquid Solution Freezing Point. Liquid collision efficiency is 1.0, temperature at 10 bars is 330 K.

| Freezing point | 0.1 x | 0.3x | 1x | 3x | 10x |
|---|---|---|---|---|---|
| 273K | 0 | 0 | 0 | 0 | 4065.6 |
| 253K | 0 | 0 | 0 | 1282.1 | 7361.3 |
| 233K | 0 | 0.6 | 242.5 | 1094.8 | 2740.0 |
| 213K | 0.6 | 22.8 | 103.3 | 354.7 | 690.1 |
| 193K | 0.3 | 3.1 | 16.2 | 65.9 | 165.9 |
| *173K* | 0.1 | 0.5 | 3.6 | 8.8 | 19.8 |
| 153K | 0.01 | 0.1 | 0.5 | 0.9 | 2.1 |

**Table 6:** Electrical energy flux above 10 bars (W m⁻²) vs. Water Abundance and Liquid Solution Freezing Point. Liquid collision efficiency is 0.8, temperature at 10 bars is 350 K.

| Freezing point | 0.1 x | 0.3x | 1x | 3x | 10x |
|---|---|---|---|---|---|
| 273K | 0 | 0 | 0 | 0 | 1602.1 |
| 253K | 0 | 0 | 0 | 586.9 | 11602.3 |
| 233K | 0 | 0.005 | 180.1 | 2810.0 | 13030.6 |
| 213K | 0.7 | 40.8 | 781.1 | 3425.2 | 14242.3 |
| 193K | 4.5 | 174.6 | 1079.1 | 4125.9 | 15766.4 |
| *173K* | 8.8 | 200.7 | 1125.1 | 4169.9 | 15796.5 |
| 153K | 11.5 | 203.5 | 1149.0 | 4203.2 | 15905.0 |

**Table 7:** Shallow Lightning Fractions (percent) vs. Water Abundance and Liquid Solution Freezing Point [a]

| Freezing point | [b]0.1 x | 0.3 x | 1x | 3x | 10x |
|---|---|---|---|---|---|
| 273K | --,--[c] | --,-- | --,-- | --,-- | 13.1, 58.2 |
| 253K | --,-- | --,-- | --,-- | 16.6, 88.9 | 15.1, 56.2 |
| 233K | --,-- | 34.5,57.4 | 32.9, 73.9 | 22.5, 46.1 | 18.2, 40.1 |
| 213K | 50.4,38.0 | 53.3,49.0 | 25.7, 40.2 | 17.7, 37.5 | 11.9, 35.8 |
| 193K | 20.7,24.9 | 12.3,32.5 | 8.5, 35.6 | 8.3, 35.5 | 6.6, 35.5 |
| 173K | 18.7,24.4 | 11.1,32.3 | 6.8, 35.3 | 6.0, 35.0 | 4.2, 34.4 |
| 153K | 18.6,24.4 | 10.7,32.1 | 6.6, 35.3 | 5.6, 34.9 | 3.8, 34.1 |

[a] For 330 K temperature at 10 bar Liquid collision efficiency is 0.8. First number is % of lightning from 0-2 bar relative to total in 0-4 bar. Second number is % of lightning from 0-4 bar relative to total in 0-10bar. [b]0.3x is 0.3 times solar water, etc. [c] -- indicates no lightning

**Table 8:** Shallow Lightning Fractions (percent) vs. Water Abundance and Liquid Solution Freezing Point [a]

| Freezing point | [b]0.1 x | 0.3 x | 1x | 3x | 10x |
|---|---|---|---|---|---|
| 273K | --,--[c] | --,-- | --,-- | --,-- | 15.1,90.4 |
| 253K | --,-- | --,-- | --,-- | 29.8,89.1 | 22.8,61.6 |
| 233K | --,-- | 0.0,15.1 | 58.8,73.1 | 37.4,47.6 | 24.9,49.2 |
| 213K | 13.8,20.0 | 68.3,45.1 | 31.7,36.4 | 24.9,38.9 | 15.0,46.4 |
| 193K | 38.5,26.0 | 34.4,29.1 | 16.0,31.8 | 14.9,36.5 | 8.3,45.1 |
| 173K | 34.1,24.6 | 33.6,28.8 | 14.5,31.4 | 13.6,36.1 | 7.5,44.9 |

| | 153K | 31.5,23.9 | 33.5,28.8 | 14.0,31.2 | 13.1,36.0 | 7.1,44.7 |

[a] *For 350 K temperature at 10 bar. Liquid collision efficiency is 0.8.. First number is % of lightning from 0-2 bar relative to total in 0-4 bar.*
*Second number is % of lightning from 0-4 bar relative to total in 0-10bar.*
[b]*0.3x is 0.3 times solar water, etc.* [c]*- - indicates no lightning*

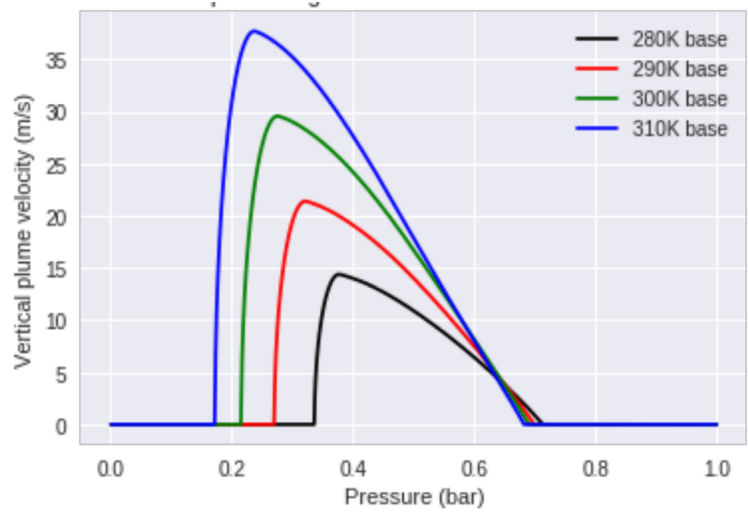

(a)

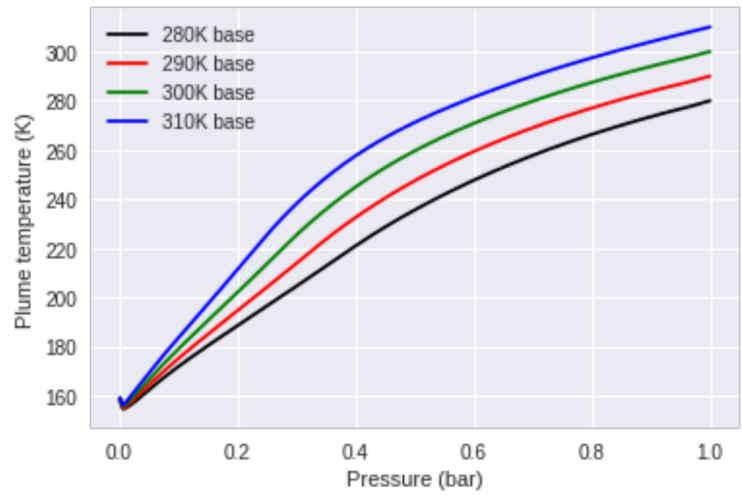

(b)

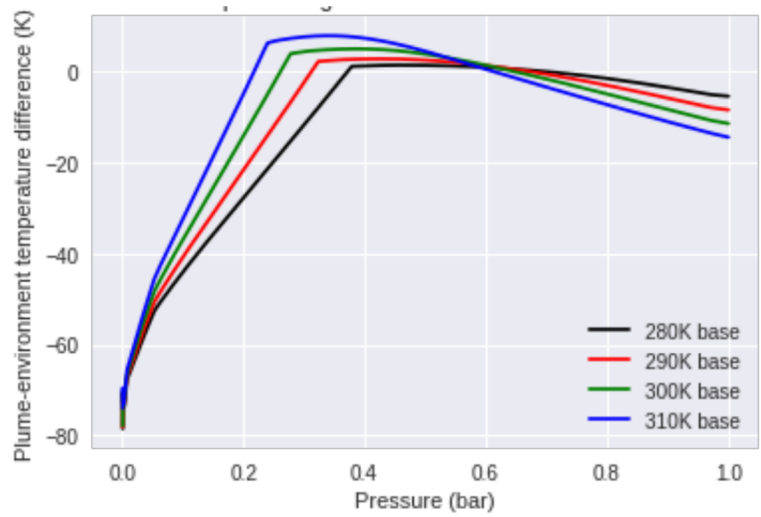

(c)

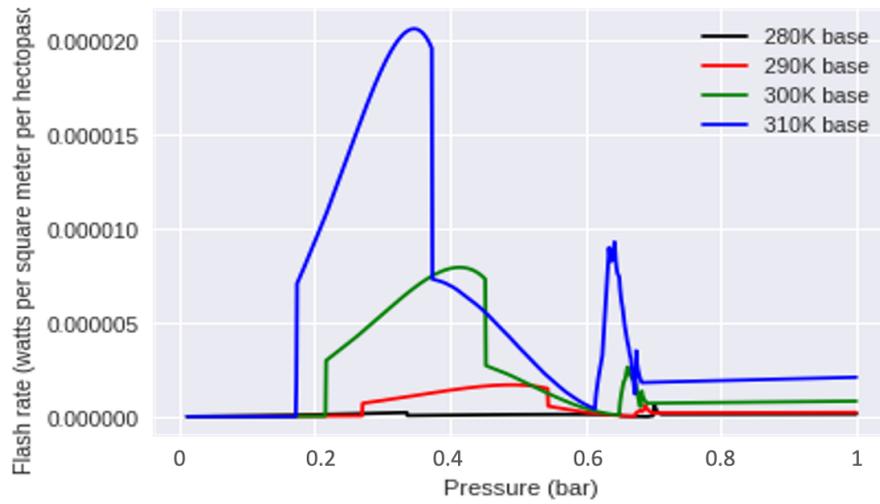

(d)

*Figure 1: Earth profiles of (top to bottom) (a) updraft velocity, (b) plume temperature, (c)plume-environment temperature difference, and (d) flash rate for freezing point 253 K and liquid collision efficiency 0.8.*

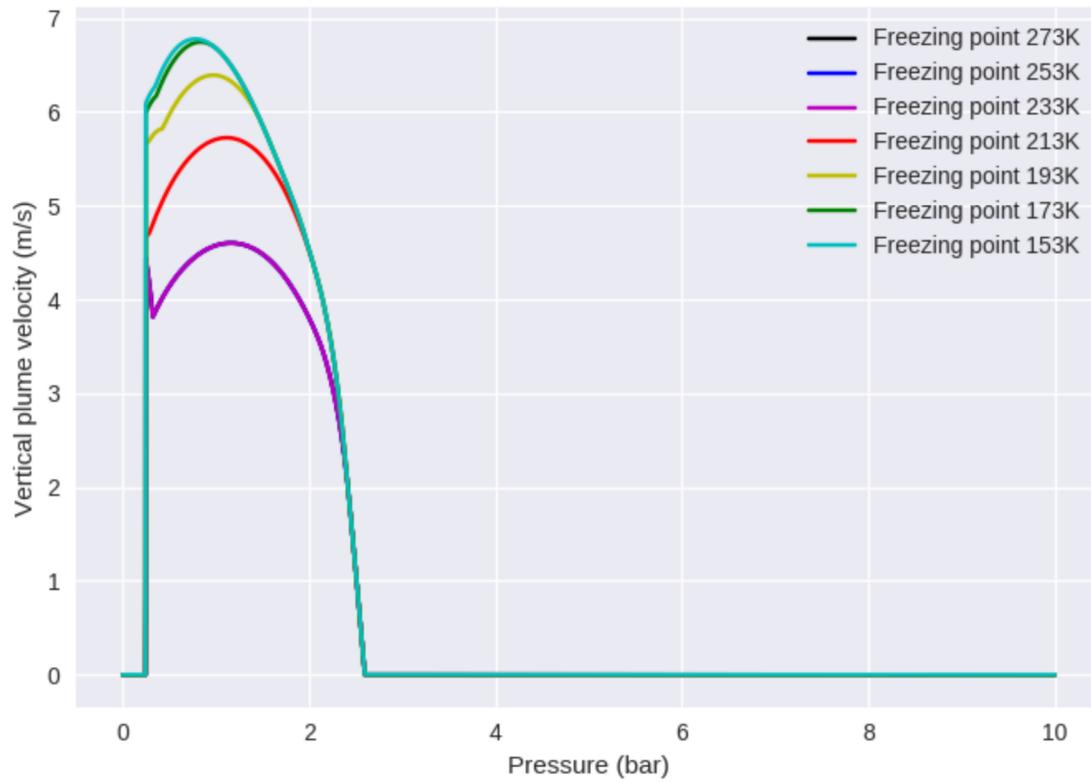

*(a)*

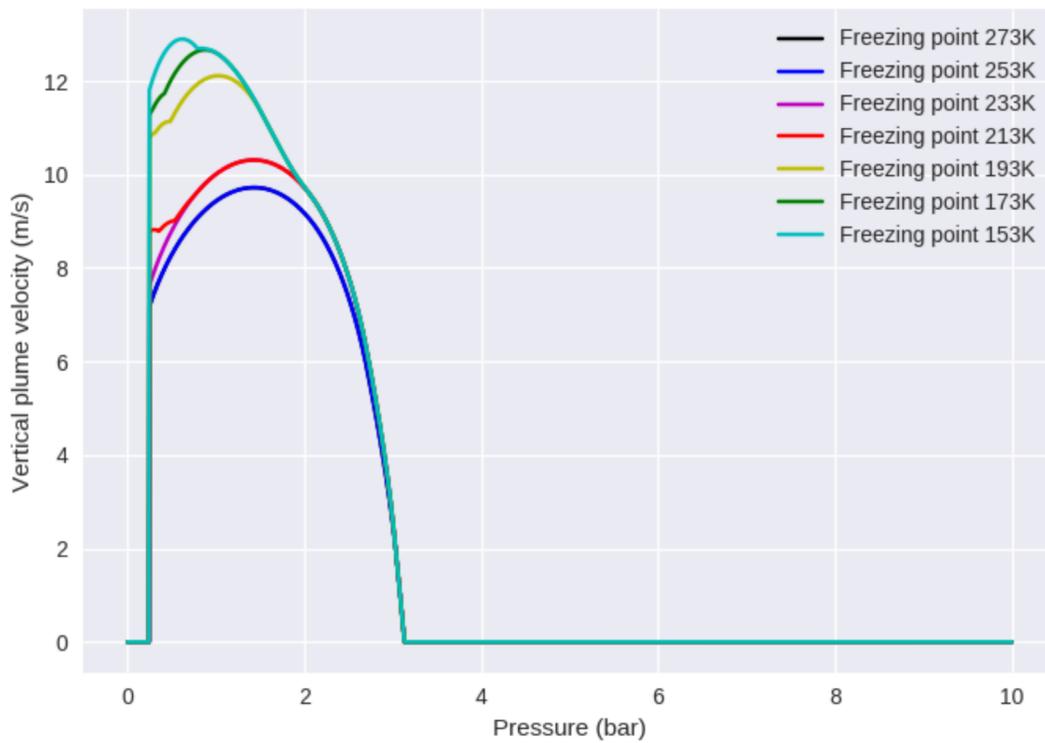

*(b)*

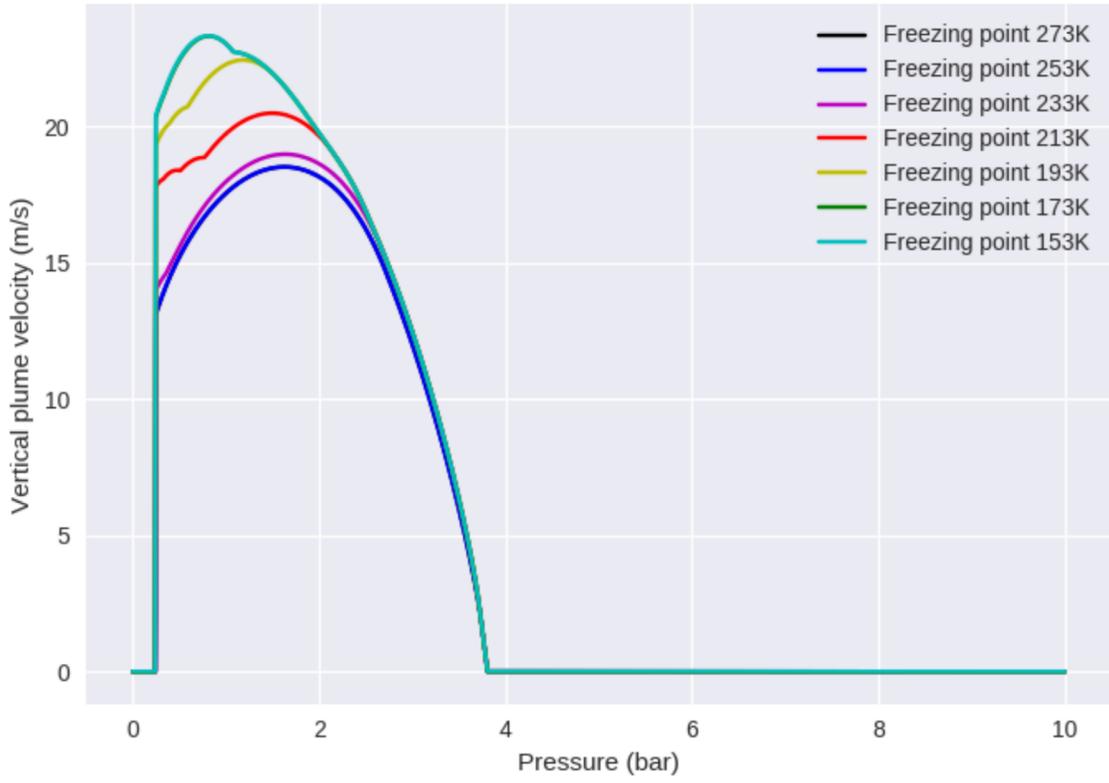

*(c)*

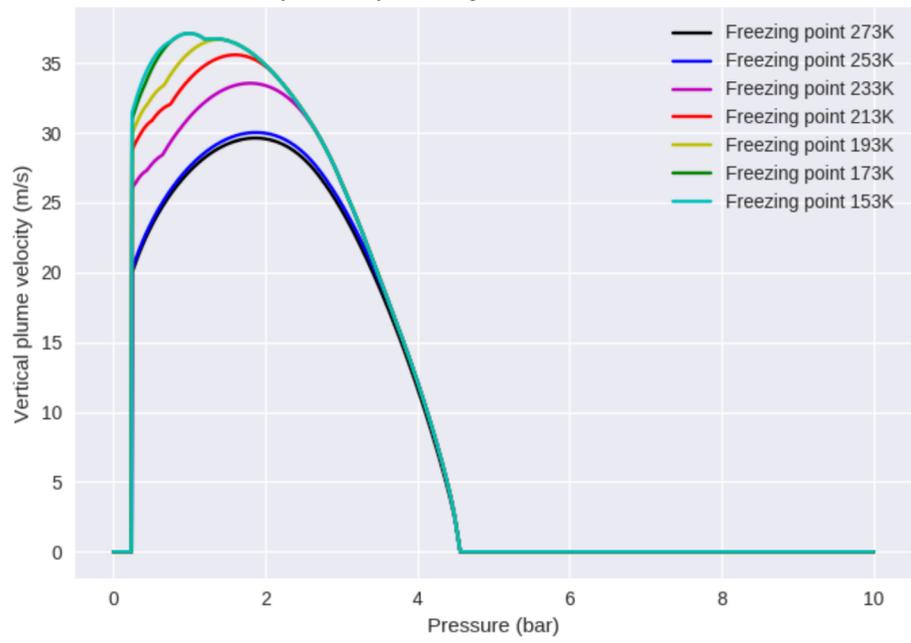

*(d)*

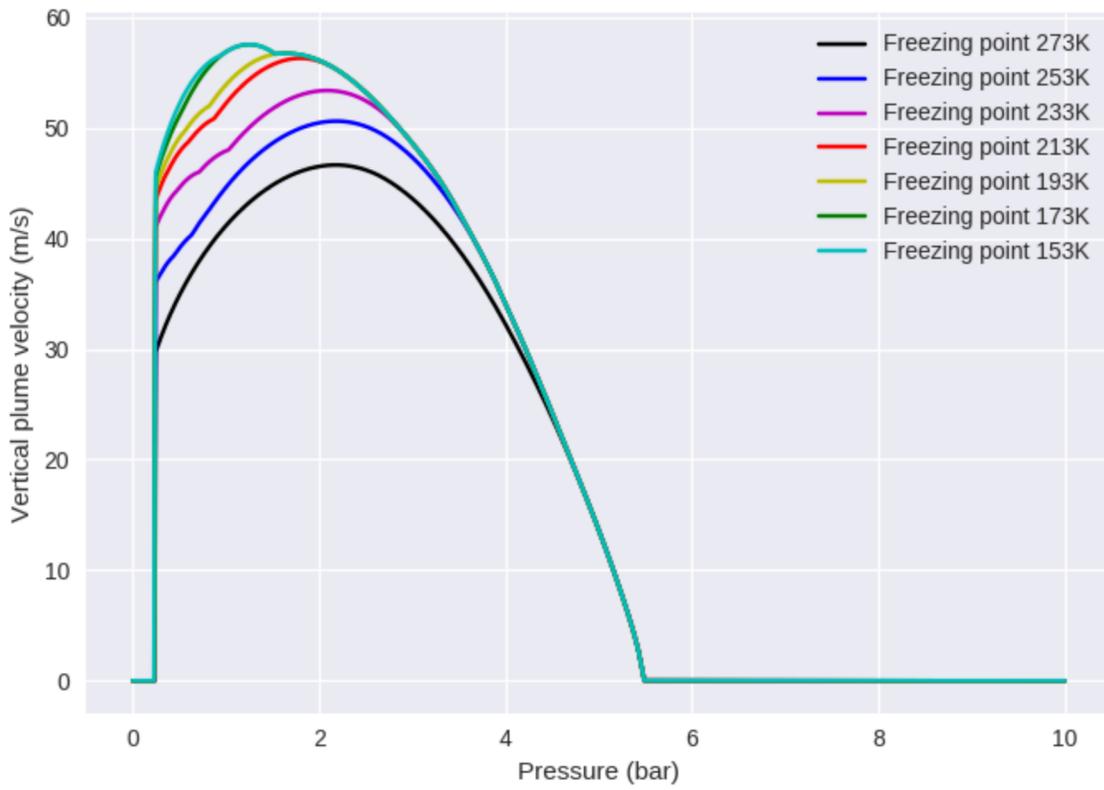

*(e)*

*Figure 2: Velocity profiles for different choices of liquid solution freezing point; from top to bottom, (a) 0.1x, (b) 0.3x, (c) 1x, (d) 3x, and (e) 10x solar water. (Environment temperature 330 K at 10 bars.)*

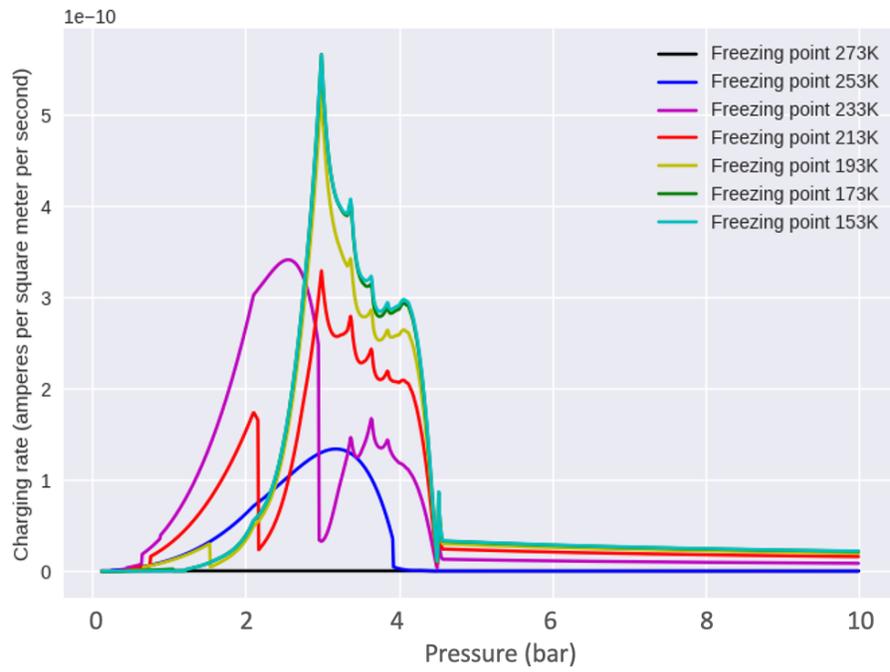

(a)

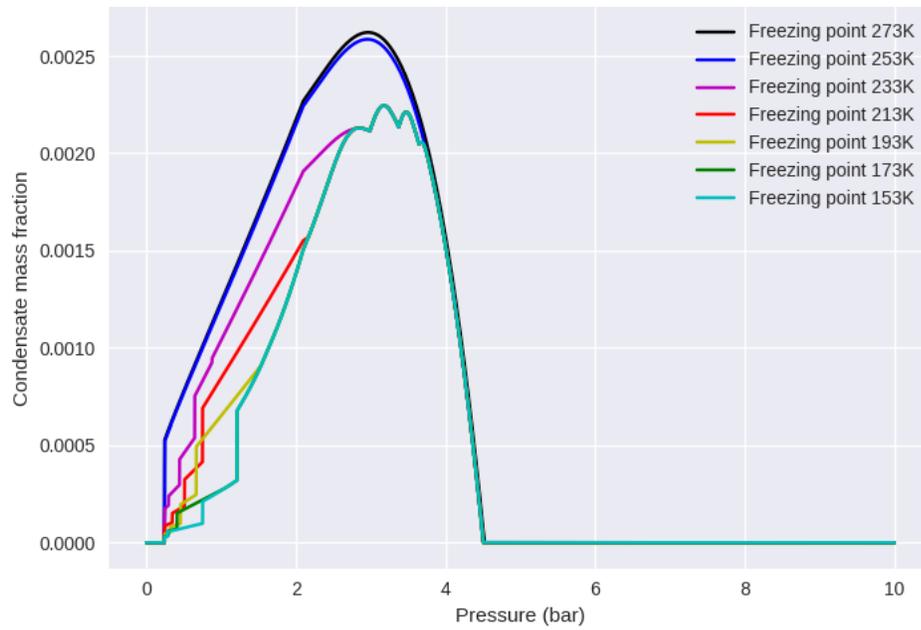

(b)

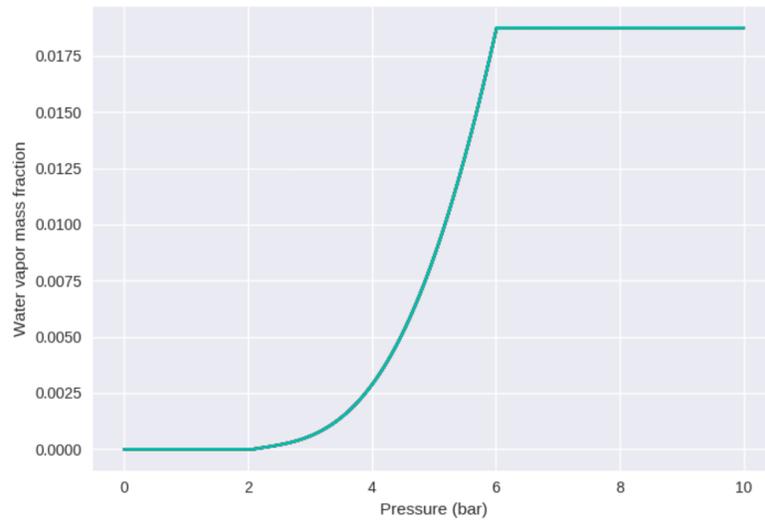

*(c)*

Figure 3: (a) Charging rate, (b) Condensate mass fraction, and (c) water vapor mass fraction . (Environmental temperature 330 K at 10 bars.)

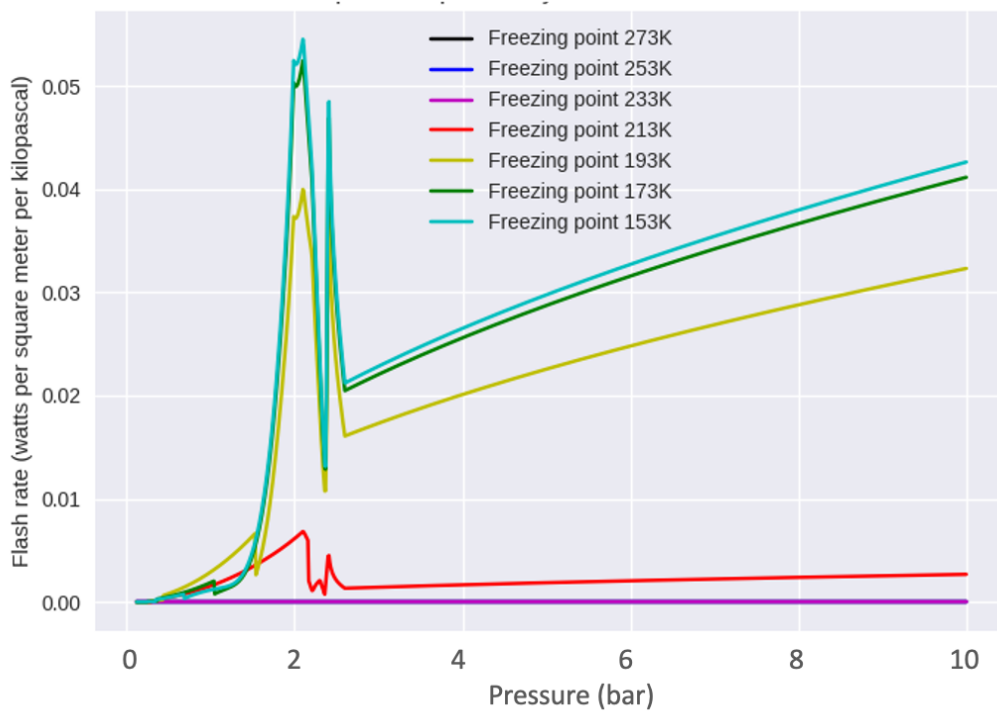

(a)

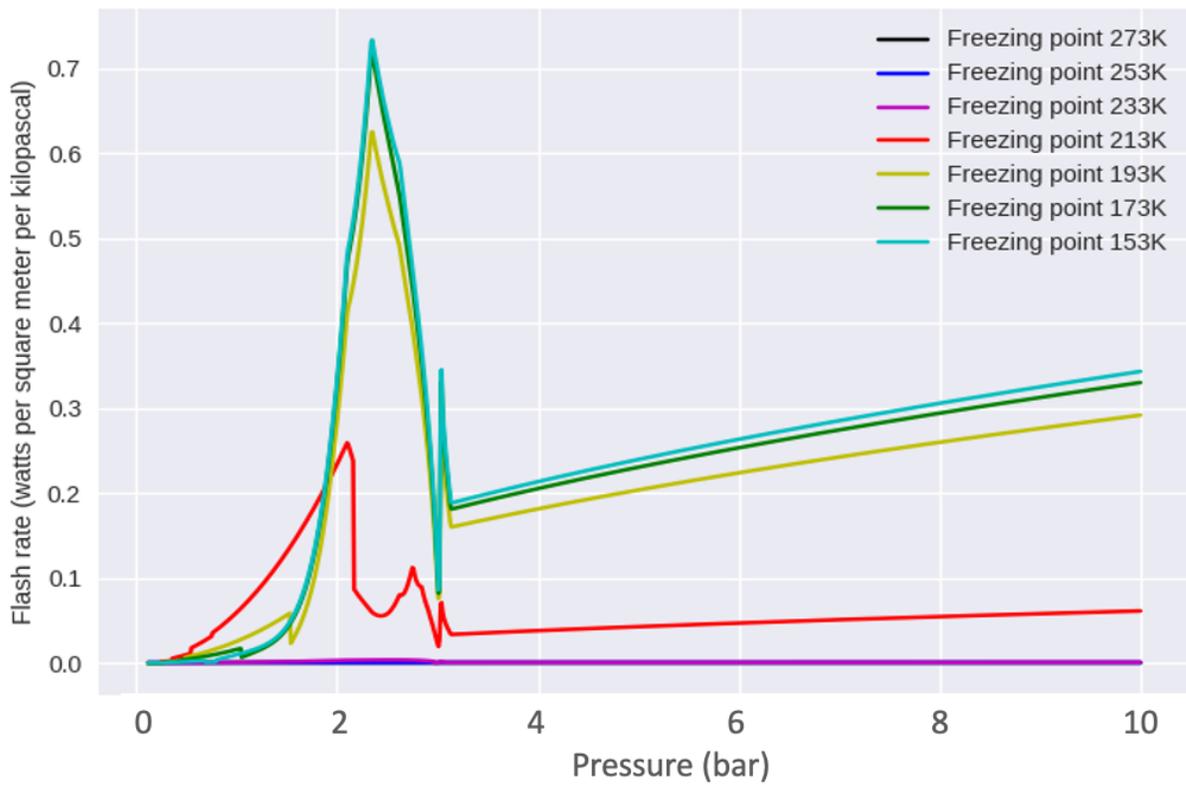

(b)

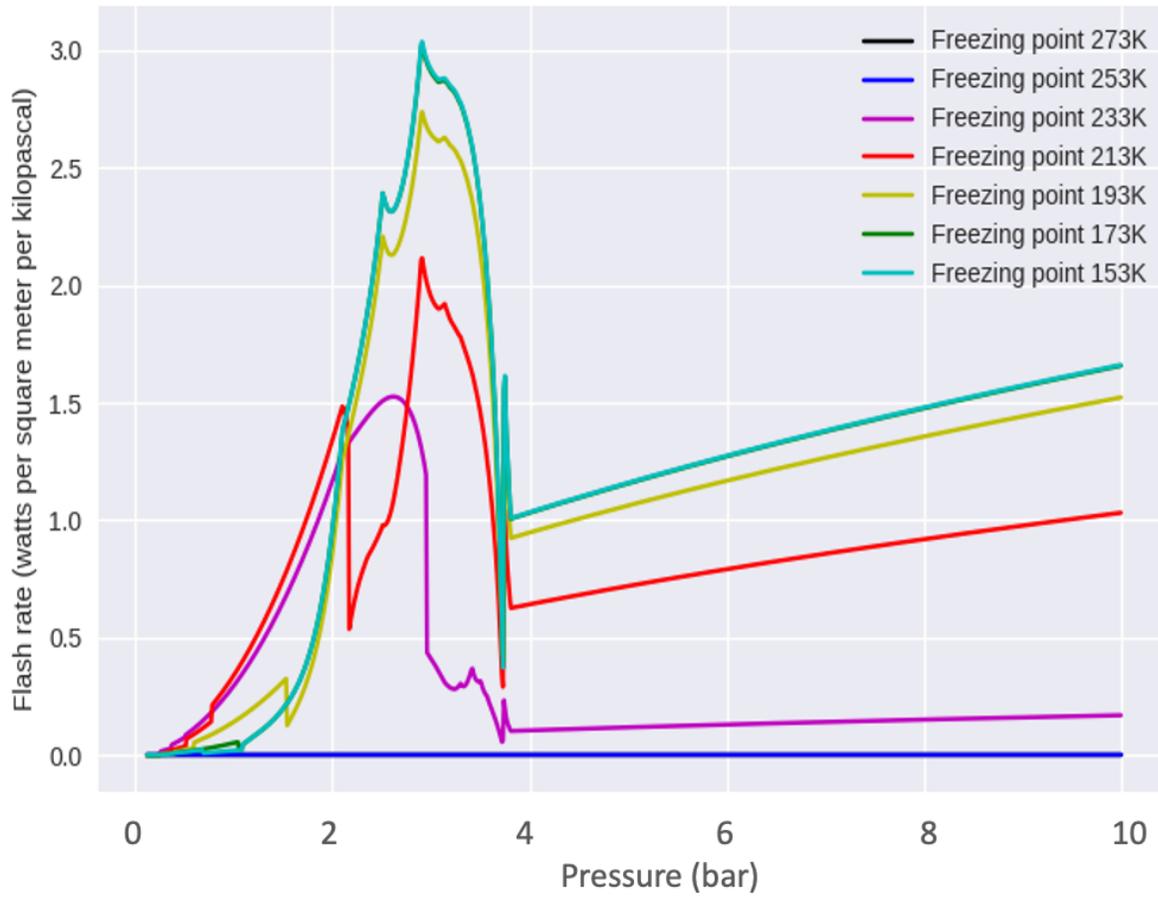

(c)

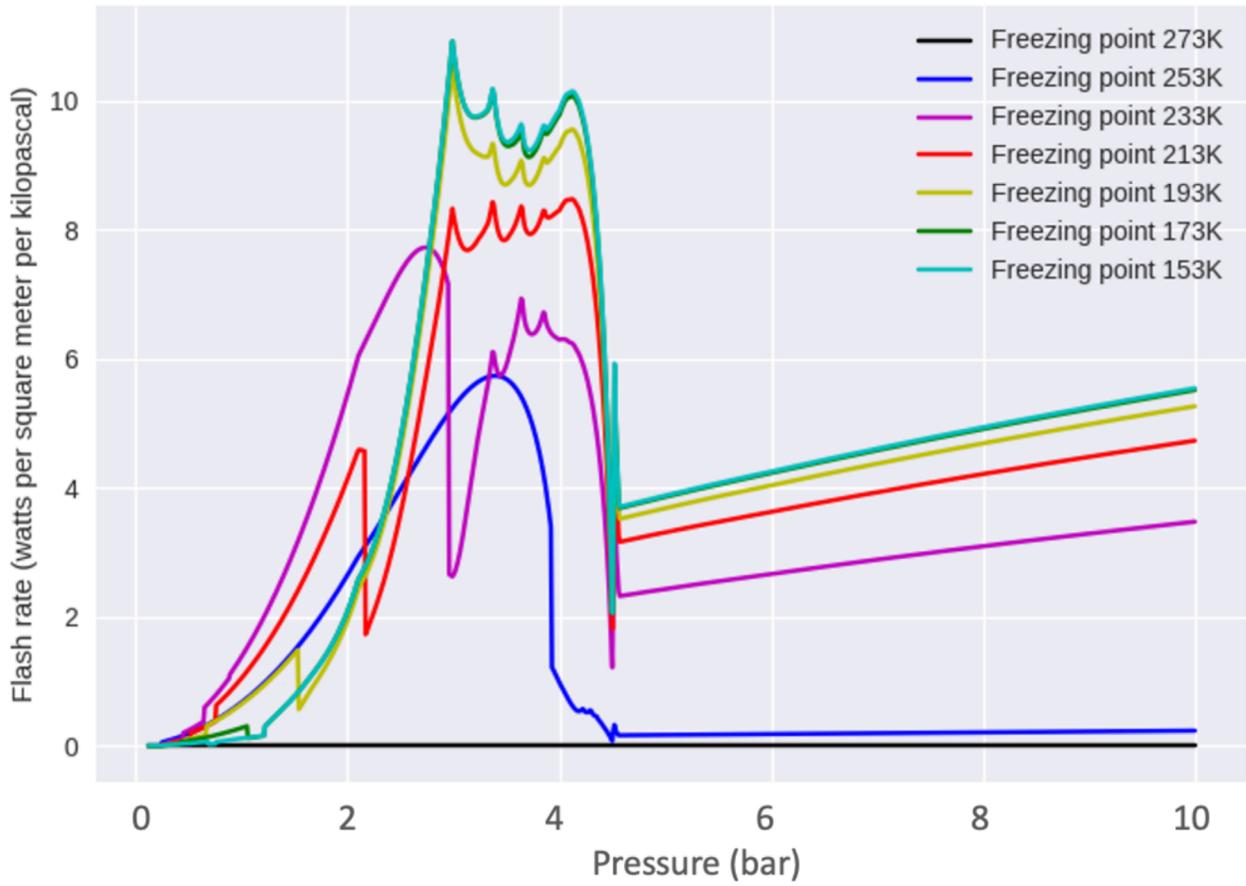

(d)

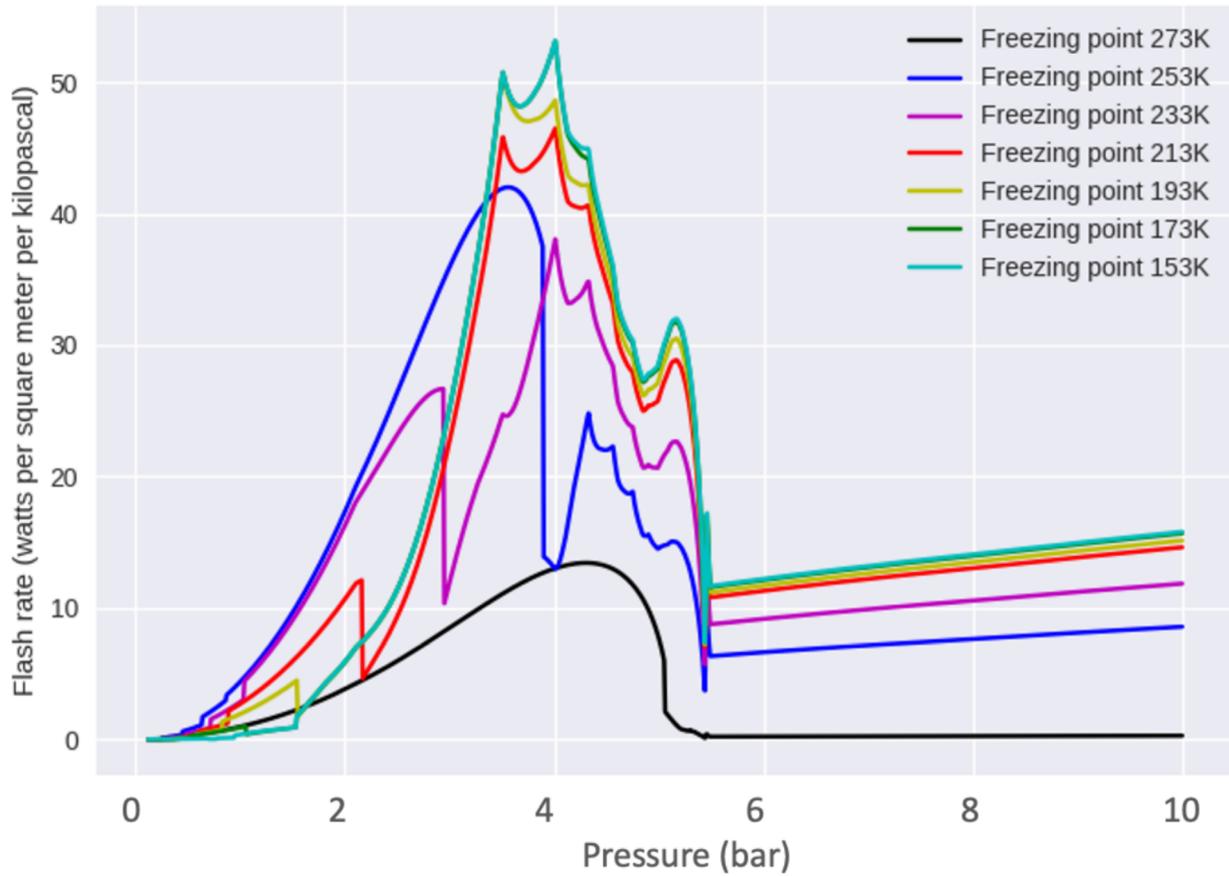

(e)

*Figure 4: Flash rate profiles for different choices of liquid solution freezing point; from top to bottom, (a) 0.1x, (b) 0.3x, (c) 1x, (d) 3x, and (e) 10x solar water. (Environment temperature 330 K at 10 bars.)*

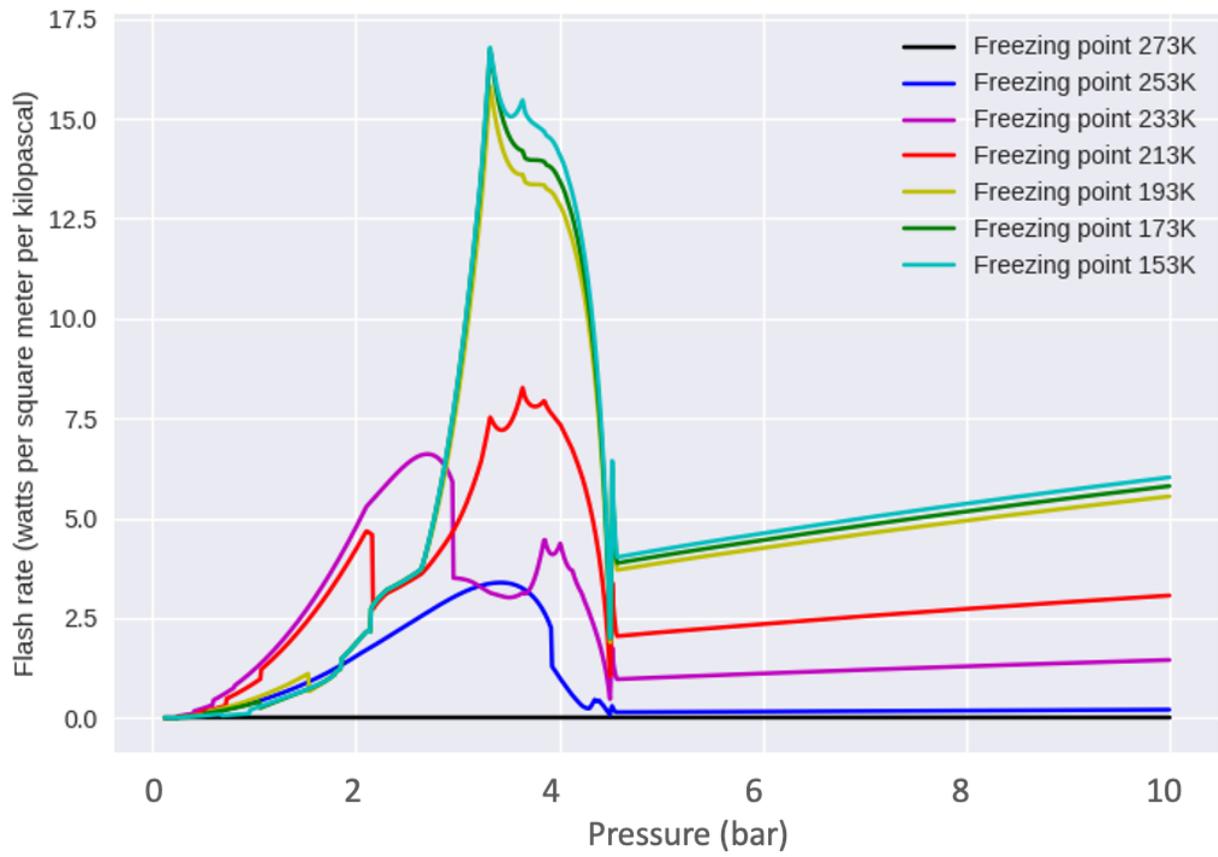

*(a)*

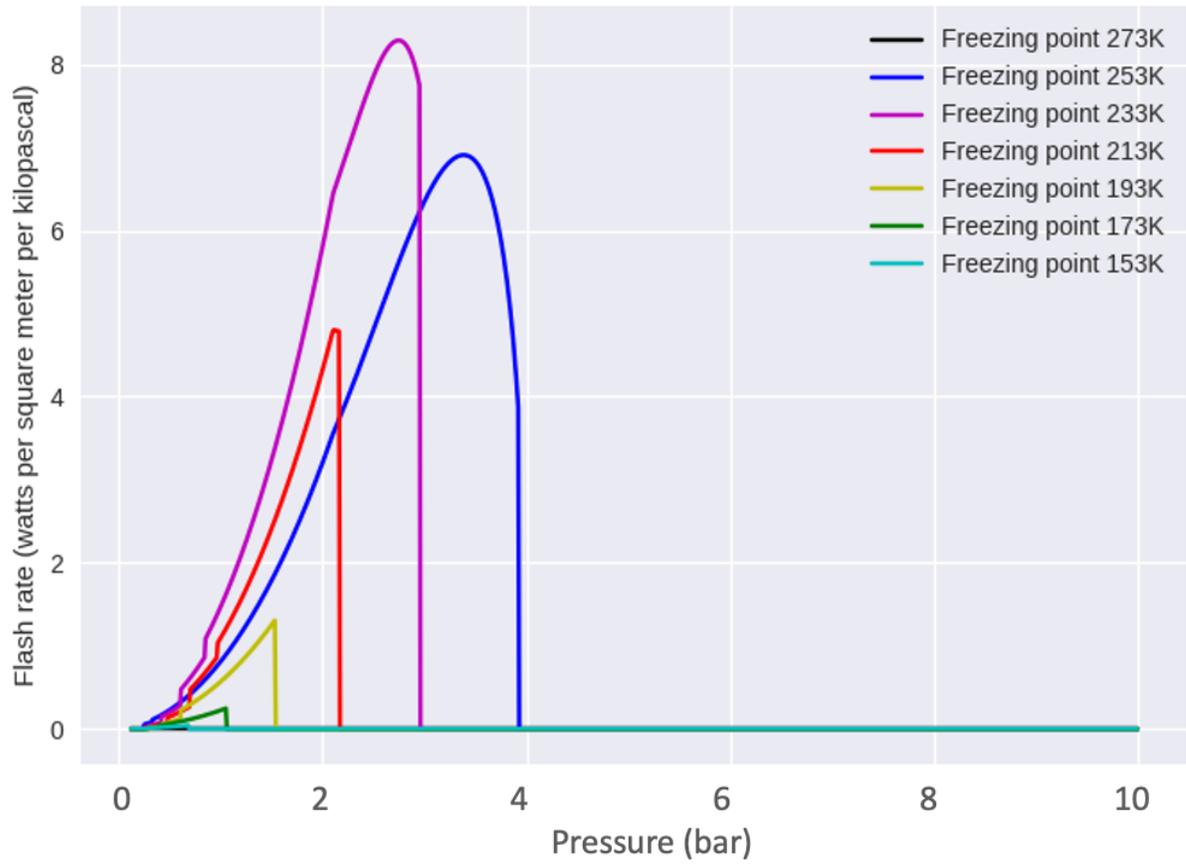

*(b)*

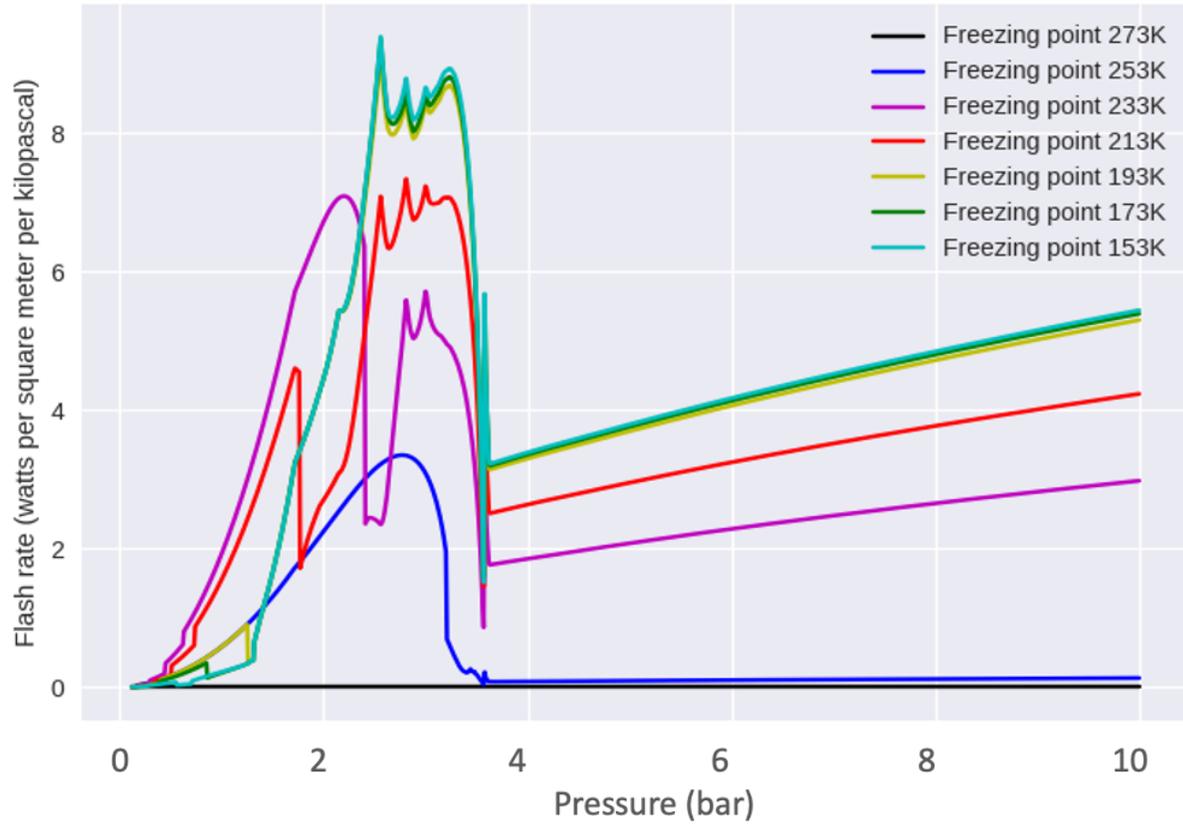

*(c)*

*Figure 5: Flash rate profiles for alternate choices of temperature profile and liquid collision efficiency; from top to bottom, (a) 330K and 0.5, (b) 330K and 1.0, (c) 350K and 0.8. (3x solar water.)*

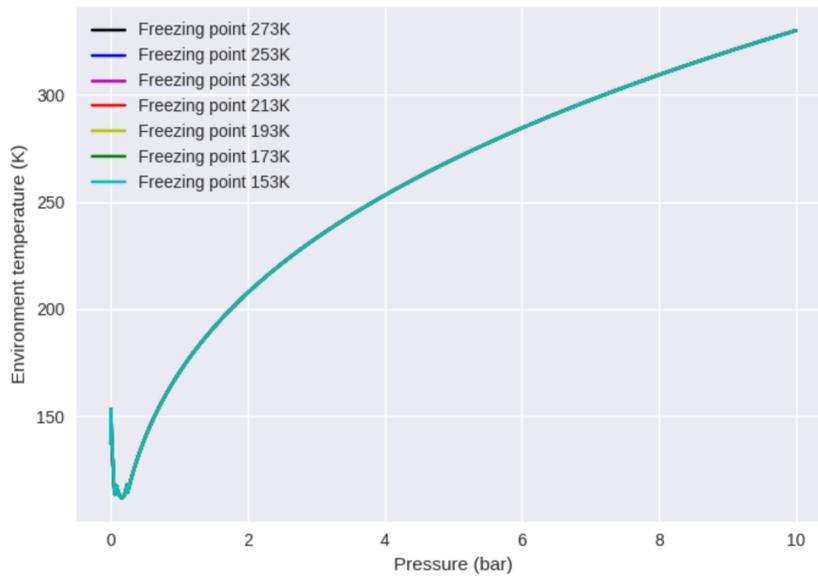

(a)

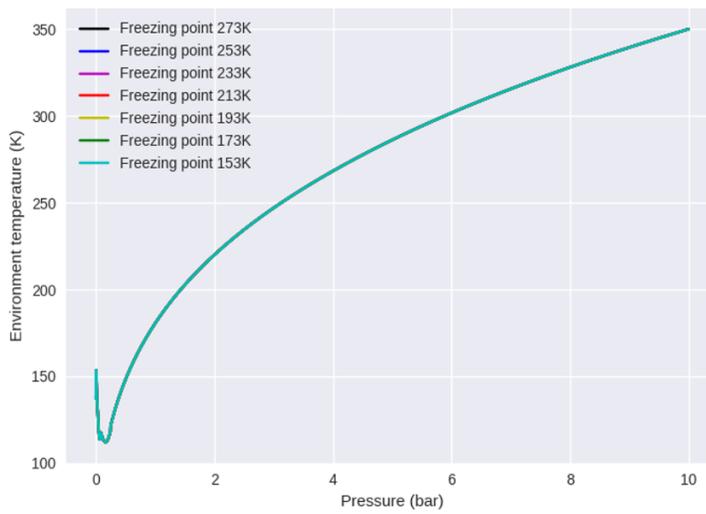

*(b)*

*Figure 6: Environment temperature profile; (a) 330 K at 10 bars profile, (b) 350 K at 10 bars profile*

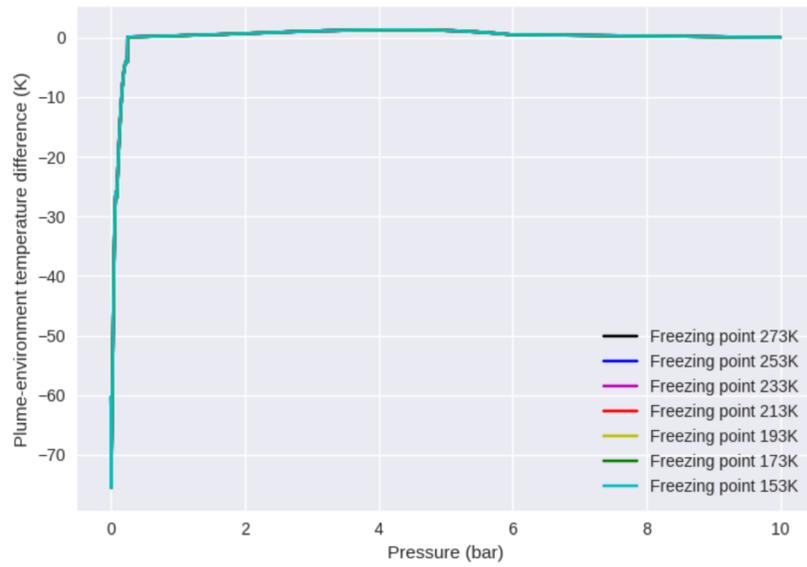

*Figure 7: Profile of plume-environment temperature difference. (3x solar water, environment temperature 330 K at 10 bars.)*